\documentclass[fleqn,usenatbib]{mnras}

\usepackage{newtxtext,newtxmath}
\usepackage[T1]{fontenc}
\usepackage{ae,aecompl}

\usepackage{graphicx}	% Including figure files
\usepackage{dblfloatfix}
\usepackage{amsmath}	% Advanced maths commands
\usepackage{threeparttable}
\usepackage{subcaption}
\usepackage{caption}
\usepackage{color}
\usepackage[dvipsnames]{xcolor}
\usepackage[normalem]{ulem}
\usepackage{booktabs}

\newcommand{\fermi}{\textit{Fermi}-{\rm LAT}}
\newcommand{\planck}{\textit{Planck}}

\newcommand{\gray}{$\gamma$-ray}
\newcommand{\grays}{$\gamma$-rays}
\newcommand{\xray}{$\rm X$-ray}

\newcommand{\msun}{\mbox{$M_\odot$}}

\def\deg{\hbox{$^\circ$}}

\title[\fermi\ detection towards RSGC 1]{Diffuse \gray\ emission from the vicinity of young massive star cluster RSGC 1}
\author[Sun et.al]{Xiao-Na Sun$^{1}$$^{2}$
Rui-Zhi Yang$^{3}$$^{4}$$^{5}$
Xiang-Yu Wang$^{1}$$^{2}$
\\
$^{1}$School of Astronomy and Space Science, Nanjing
University, Nanjing 210093, China\\
$^{2}$Key laboratory of Modern Astronomy and Astrophysics, Nanjing University, Ministry of Education, Nanjing 210093, China\\
$^{3}$Department of Astronomy, School of Physical Sciences, University of Science and Technology of China, Hefei, Anhui 230026, China\\
$^{4}$CAS Key Labrotory for Research in Galaxies and Cosmology, University of Science and Technology of China, Hefei, Anhui 230026, China\\
$^{5}$School of Astronomy and Space Science, University of Science and Technology of China, Hefei, Anhui 230026, China\\
}

\pubyear{2020}

\begin{document}
\label{firstpage}
\pagerange{\pageref{firstpage}--\pageref{lastpage}}
\maketitle

\begin{abstract}
We report the Fermi Large Area Telescope (\fermi) detection of the \gray\ emission towards the young massive star cluster RSGC 1.
Using the latest source catalog and diffuse background models, we found that the diffuse \gray\ emission in this region can be resolved into three different components. 
The GeV \gray\ emission from the region HESS J1837-069 has a photon index of $1.83\pm0.08$. Combining with the HESS and MAGIC data, we argue that the \gray\ emission in this region  likely originate from a pulsar wind nebula (PWN).
The \gray\ emission from the northwest part (region A) can be modelled by an ellipse with the semimajor and semiminor axis of $0.5^{\circ}$ and $0.25^{\circ}$, respectively. The GeV emission has a hard spectrum with a photon index of $2.05\pm 0.02$ and is partially coincide with the TeV source MAGIC J1835-069. The possible origin of the \gray\ emission in this region is the interaction of the cosmic rays (CRs) accelerated by SNR G24.7+0.6 or/and the OB cluster G25.18+0.26 with the surrounding gas clouds. 
The GeV \gray\ emission from the southeast region (region B) can be modeled as an ellipse with the semimajor and semiminor axis of $0.9^{\circ}$ and $0.5^{\circ}$, respectively, and also reveals a hard \gray\ spectrum. We argue that the most probable origin is the interaction of the  accelerated protons in the young massive star cluster RSGC 1  with ambient gas clouds, and the total cosmic-ray (CR) proton energy is estimated to be as high as  $\sim 1\times10^{50}\ \rm erg$. 
\end{abstract}


%%%%%%%%%%%%%%%%%%%%%%%%%%%%%%%%%%%%%%%%%%%%%%%%%%

%%%%%%%%%%%%%%%%% BODY OF PAPER %%%%%%%%%%%%%%%%%%

%%%%%%%%%%%%%%%%% introduction %%%%%%%%%%%%%%%%%%
\section{Introduction}
The origin of CRs is still a puzzle. 
There is a consensus in the CR community that the main acceleration sites of Galactic CRs are supernova remnants (SNRs) \citep{Drury12,Blasi13}. However, recently there are growing evidences that the young massive star clusters can also play an important role in accelerating  the CRs in the Galaxy \citep{Ackermann11, Aharonian19, Katsuta17}. In several such systems, e.g., Cygnus cocoon \citep{Ackermann11, Aharonian19}, Westerlund 1 \citep{Abramowski12}, Westerlund 2 \citep{Yang18}, NGC 3603 \citep{Yang17} and 30 Dor C \citep{Abramowski15}, the extended \gray\ (from GeV to TeV) emissions with hard spectra have been detected.

In this regard, RSGC 1 is one of the most massive young  star clusters \citep[see, e.g.,][]{zwart10,davies12} with an initial mass estimated to be $2 \times 10^{4} - 4 \times 10^{4}$\msun\ and an age of $\sim 10 \rm~Myr$ \citep{Figer06,davies12}.
It is one of the rare clusters in the Galaxy containing more than 10  red super-giants (RSGs) \citep{Figer06,davies12}.
HESS telescope array found the diffuse \gray\ source HESS J1837-069 close to RSGC 1 \citep{Aharonian05,Aharonian06},  but \citet{Fujita14} argued that the TeV emission should be connected with the PWN. Two \xray\ PWN candidates, AX J1837.3-0652 and AX J1838.0-0655, adjacent to HESS J1837-069 were discovered \citep{Gotthelf08}.
%----------------------------------------------------- FIGURE 1
\begin{figure*}
%\centering
\includegraphics[scale=0.35]{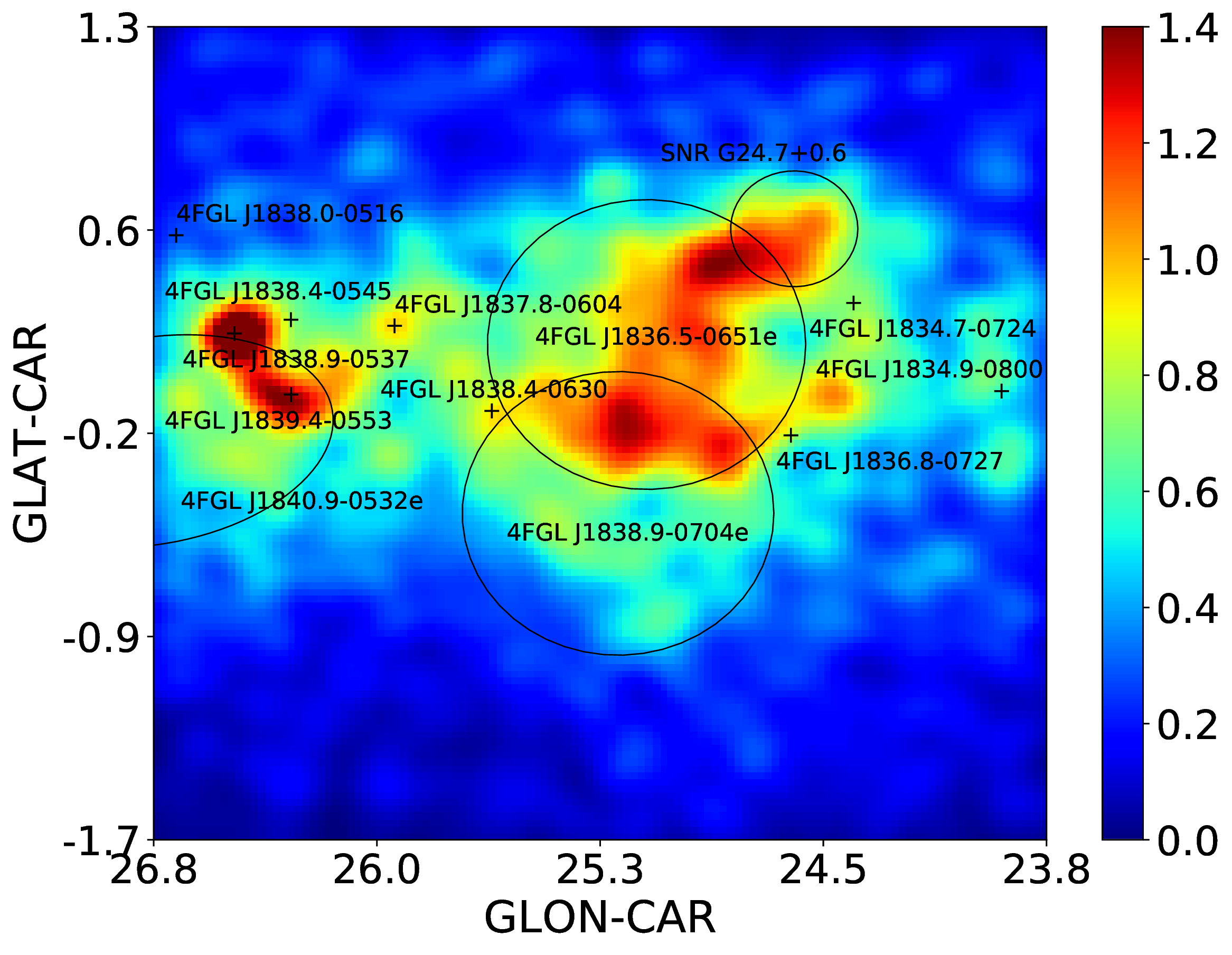}
\includegraphics[scale=0.35]{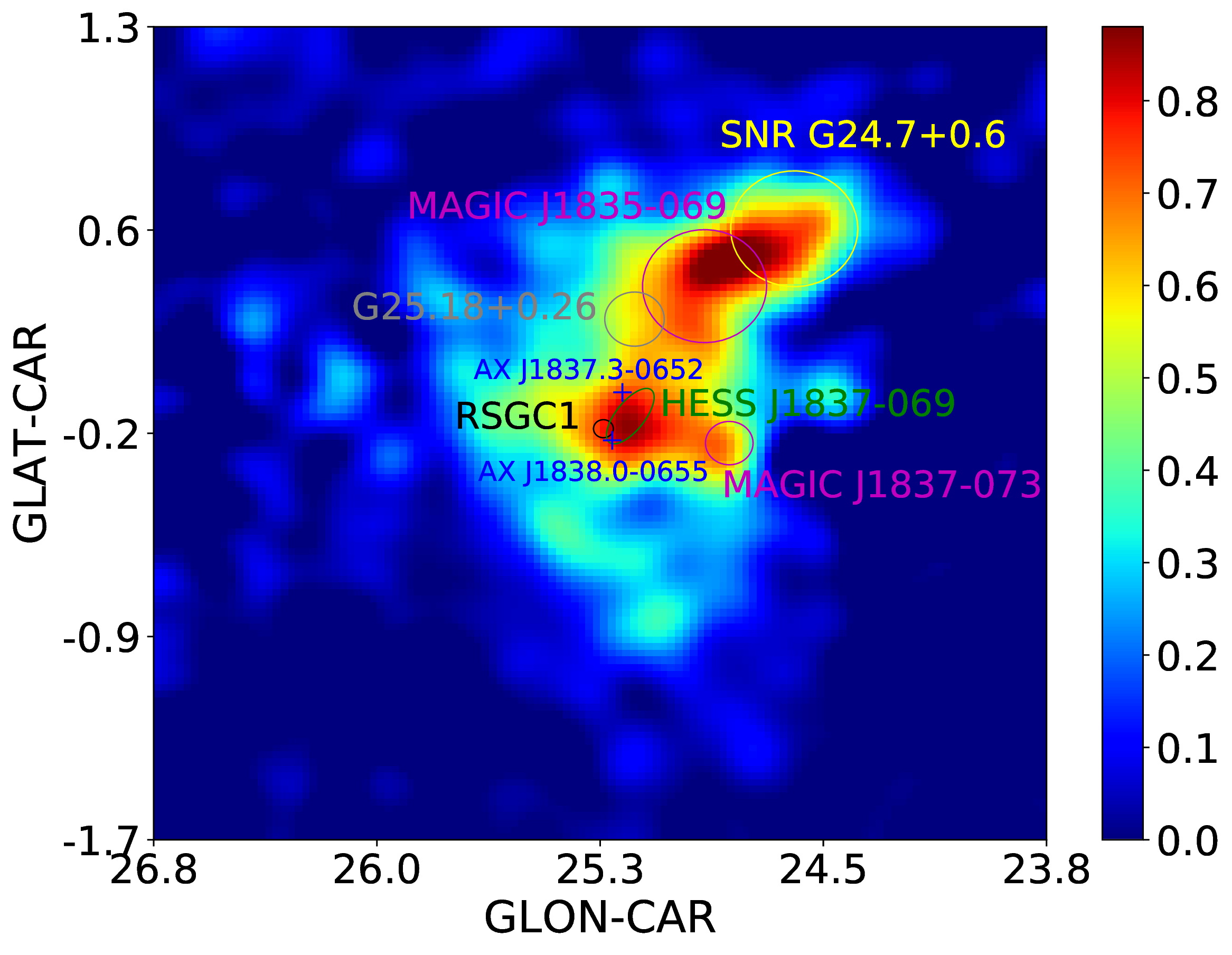}
\caption {Left: \fermi\ counts map above 10 GeV in the $3\deg \times 3\deg$ region surrounding RSGC 1, with pixel size corresponding to $0.025\deg \times 0.025\deg$, smoothed with a Gaussian filter of $0.23\deg$. All black marks represent the 4FGL sources within the region.
Right: Residual map above 10 GeV near the RSGC 1 complex after subtracting the background sources. The marks with different colors show the signals detected by multiwavelength surveys, details on the context are given in Sect.~\ref{sec:spatial_analy}.}
\label{fig:cmap}
\end{figure*}
\cite{Lande12} first found the \fermi\ \gray\ extension in the $1.7\deg \times 2.1\deg$ region around $(l, b) = (25.0\deg, 0.0\deg)$ (G25 region), which is spatially coincident with the TeV source HESS J1837-069.
RSGC 1 is just located within the complex G25 region in projection on the sky \citep{Fujita14, Katsuta17}.
\citet{Katsuta17} re-analyzed this complex region using almost five years of \fermi\ data and proposed that the extended GeV emissions (G25A and G25B) are possibly associated with a massive OB association G25.18+0.26.
Recently \citet{MAGIC19}  reported the TeV observations towards this region using MAGIC telescopes.  They indicated that a new TeV source MAGIC J1835-069, which is likely associated with the SNR G24.7+0.6 \citep{Reich84,Leahy89}, and the detected \gray\ emission can be interpreted as the result of proton-proton interaction between the CRs accelerated by SNR and the CO-rich surrounding.
In addition, another TeV source MAGIC J1837-073 was also identified in this region \citep{MAGIC19}. It has a single power-law photon spectrum extending to 3 TeV,  and was believed to be part of the scarcely young star clusters which have been confirmed to be the acceleration sites for a significant amount of the Galactic CRs \citep{MAGIC19} .
The SNR G24.7+0.6 mentioned above is a filled-center SNR discovered at radio frequencies \citep{Reich84,Leahy89} with an age of $\sim 9500~\rm yr$ \citep{Leahy89}.
\cite{Reich84} indicated that the radio emission in this region is caused by a central PWN powered by an undetected pulsar.
Based on infrared data \cite{Petriella10} suggested that the formation of some young stellar object may have been triggered in the vicinity of the SNR G24.7+0.6.
To summarize, the RSGC 1 vicinity is a complex region in \grays.\ The rich GeV/TeV emitting phenomena make it an interesting target to study the physics related with CR acceleration and the corresponding \gray\ radiation process. 

In this paper, we perform a detailed analysis on the 10-year \fermi\ data towards this region and try to pin down the \gray\ radiation mechanism therein. The paper is organized as follows. In Sec.2, we present the detail of the data analysis. In Sec.3, we study the possible gas distributions. In Sec.4, we investigate the possible radiation mechanisms of the \gray\ emissions. In Sec.5, we discuss the implications of our results. 

%%%%%%%%%%%%%%%%% data reduction and analysis %%%%%%%%%%%%%%%%%%
\section{\fermi\ data analysis}
We analyze the \fermi\ Pass 8 database around the RSGC 1 region from August 4, 2008 (MET 239557417) until August 3, 2019 (MET 586492988) with both the front and back converted photons considered.
A 7\deg $\times$ 7\deg\ square region centered at the position of RSGC 1 (R.A. = 279.492$\deg$, Dec. = -6.883$\deg$) is chosen as the region of interest (ROI).
We use the ``source'' event class, recommended for individual source analysis, and the recommended expression $\rm (DATA\_QUAL > 0) \&\& (LAT\_CONFIG == 1)$ to exclude time periods when some spacecraft event affected the data quality.
To reduce the background contamination from the Earth's albedo, only the events with zenith angles less than 100$\deg$ are included for the analysis.
We process the data through the current Fermitools from conda distribution\footnote{\url{https://github.com/fermi-lat/Fermitools-conda/}} together with the latest version of the instrument response functions (IRFs) {\it P8R3\_SOURCE\_V2}.
We use the python module that implements a maximum likelihood optimization technique for a standard binned analysis\footnote{\url{https://fermi.gsfc.nasa.gov/ssc/data/analysis/scitools/python\_tutorial.html}.}

In our background model, we include the sources in the \fermi\ eight-year catalog \citep[4FGL,][]{Fermi19} within the ROI enlarged by 5$\deg$.
We leave the normalizations and spectral indices free for all sources within $6\deg$ away from RSGC 1.
For the diffuse background components, we use the latest Galactic diffuse model {\it gll\_iem\_v07.fits} and isotropic emission model {\it iso\_P8R3\_SOURCE\_V2\_v1.txt}\footnote{\url{https://fermi.gsfc.nasa.gov/ssc/data/access/lat/BackgroundModels.html}} with their normalization parameters free.

%----------------------------------------------------- TABLE 1
\begin{table*}
%\centering
        \caption{The sources around RSGC 1 \& G25. HESS J1837-069: with an orientation angle of the semimajor axis of $121 \pm 10$ degrees clockwise with respect to the positive Galactic longitude axis.}
\begin{tabular}{lcccc}
\hline
\hline
        Source & Extension type& Center position & size \\
         && ({\it l,b}) & (deg) \\
\hline
        RSGC 1& circle&(25.27, -0.15)& 0.03\\ %\cline{1-1}
        HESS J1837-069&ellipse& (25.18, -0.11) & (0.12, 0.05) \\ %\cline{1-1}
        MAGIC J1835-069&circle& (24.94, 0.37) & 0.21 \\ %\cline{1-1}
        MAGIC J1837-073&circle& (24.85, -0.21) & 0.08 \\ %\cline{1-1}
        SNR G24.7+0.6&circle& (24.63, 0.58) & 0.21 \\ %\cline{1-1}
        AX J1836.3-0647 (G25.18+0.26)&circle& (25.17, 0.25) & 0.10 \\ %\cline{1-1}
        AX J1837.3-0652& point& (25.21, -0.02) & -- \\ %\cline{1-1}
        AX J1838.0-0655& point&(25.25, -0.19) & -- \\ %\cline{1-1}
\hline
\hline
\end{tabular}
\label{table:1}
\end{table*}

\subsection{Spatial analysis}
\label{sec:spatial_analy}
We use the events above 10 GeV to study the spatial distribution of the \gray\ emission near the RSGC 1 complex. The counts map in the $3\deg \times 3\deg$ region around RSGC 1 is shown in Fig.\ref{fig:cmap}, left, with pixel size corresponding to $0.025\deg \times 0.025\deg$, smoothed with a Gaussian filter of $0.23^ \circ$. The black crosses and extended regions show the positions of the sources from 4FGL within the region.
We exclude three extended sources (SNR G24.7+0.6, 4FGL J1836.5-0651e, and 4FGL J1838.9-0704e) and two unassociated point sources (4FGL J1838.4-0630 and 4FGL J1834.7-0724) from our background model after performing the binned likelihood analysis.

The corresponding background-subtracted image is shown in the right panel of Fig.~\ref{fig:cmap}.
On the residual map, the black circle is the approximate extent of the massive star cluster RSGC 1 \citep{Figer06}.
The very high energy (VHE) \gray\ source HESS J1837-069 \citep{Aharonian05, Aharonian06} is displayed by the green ellipse. 
The extensions of MAGIC J1835-069 and MAGIC J1837-073 \citep{MAGIC19} are represented by the magenta circles.
The grey circle represents the young massive OB association/cluster G25.18+0.26 reported by \cite{Katsuta17}.
The yellow circle displays the extended region of GeV \gray\ emission from SNR G24.7+0.6 which has been listed in the 4FGL. 
The positions of two \xray\ PWN candidates (AX J1837.3-0652 and AX J1838.0-0655) in \citet{Gotthelf08} are marked with blue crosses.
Table \ref{table:1} shows the names, positions, and sizes for the eight sources.

\begin{figure}
%\centering
\includegraphics[scale=0.35]{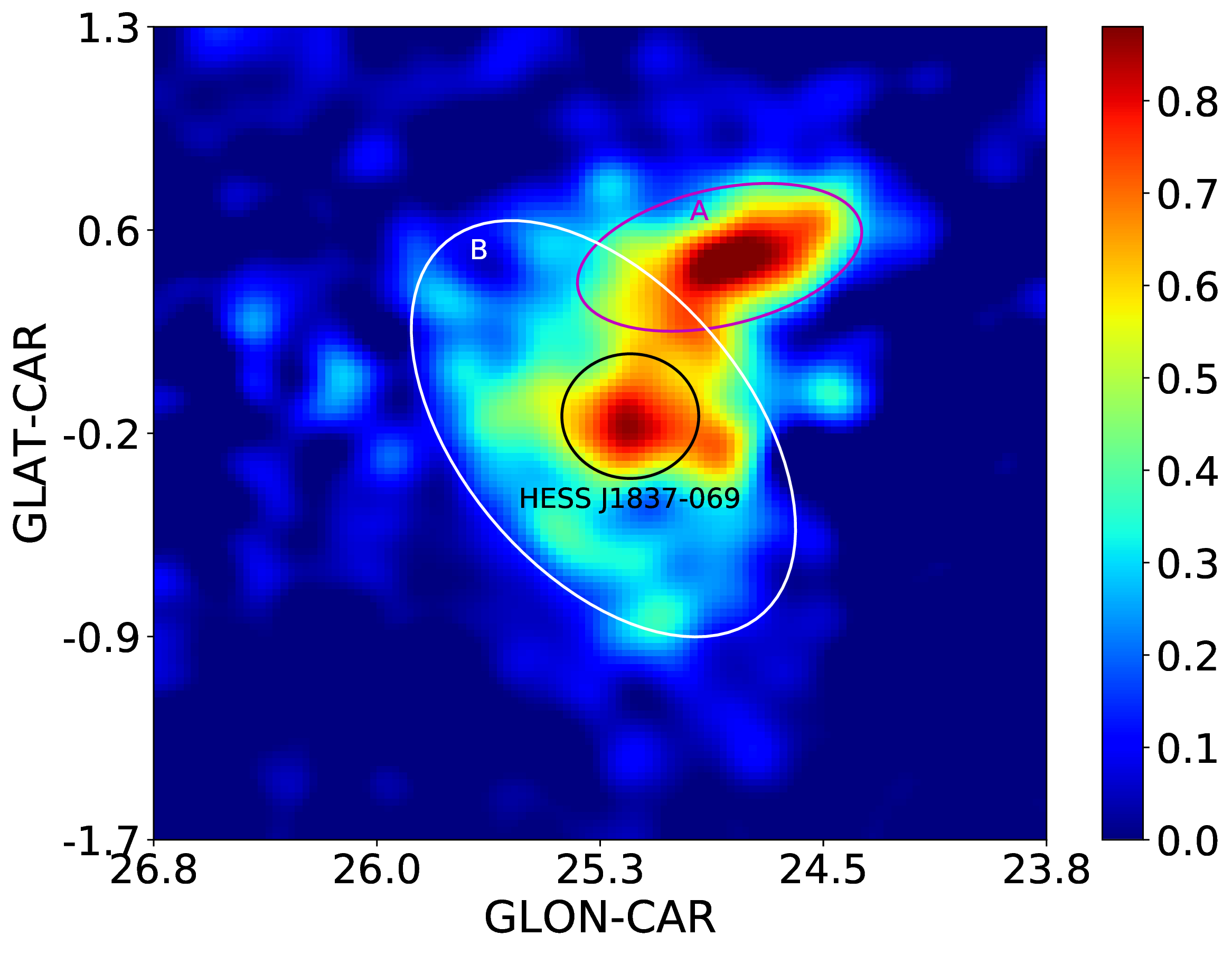}
\caption {Same residual map as Figure~\ref{fig:cmap}, right panel.
The marks represent the assumed components for spatial analysis with the maximum likelihood value.
The spatial template is constitute of A, B, and HESS J1837-069.
For details, see Sect.~\ref{sec:spatial_analy}.
}
\label{fig:spatialT}
\end{figure}

\begin{figure*}
%\centering
\includegraphics[scale=0.45]{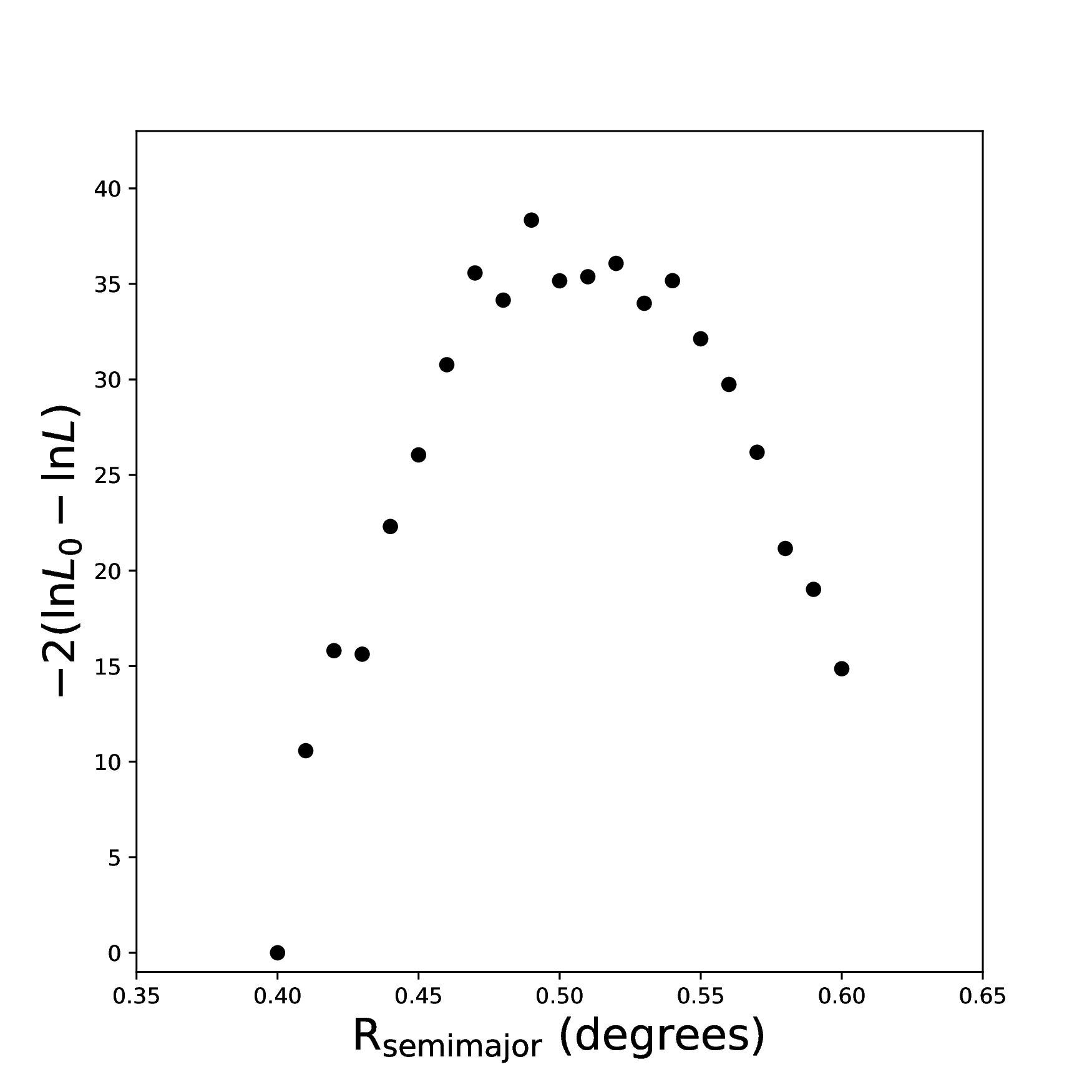}
\includegraphics[scale=0.45]{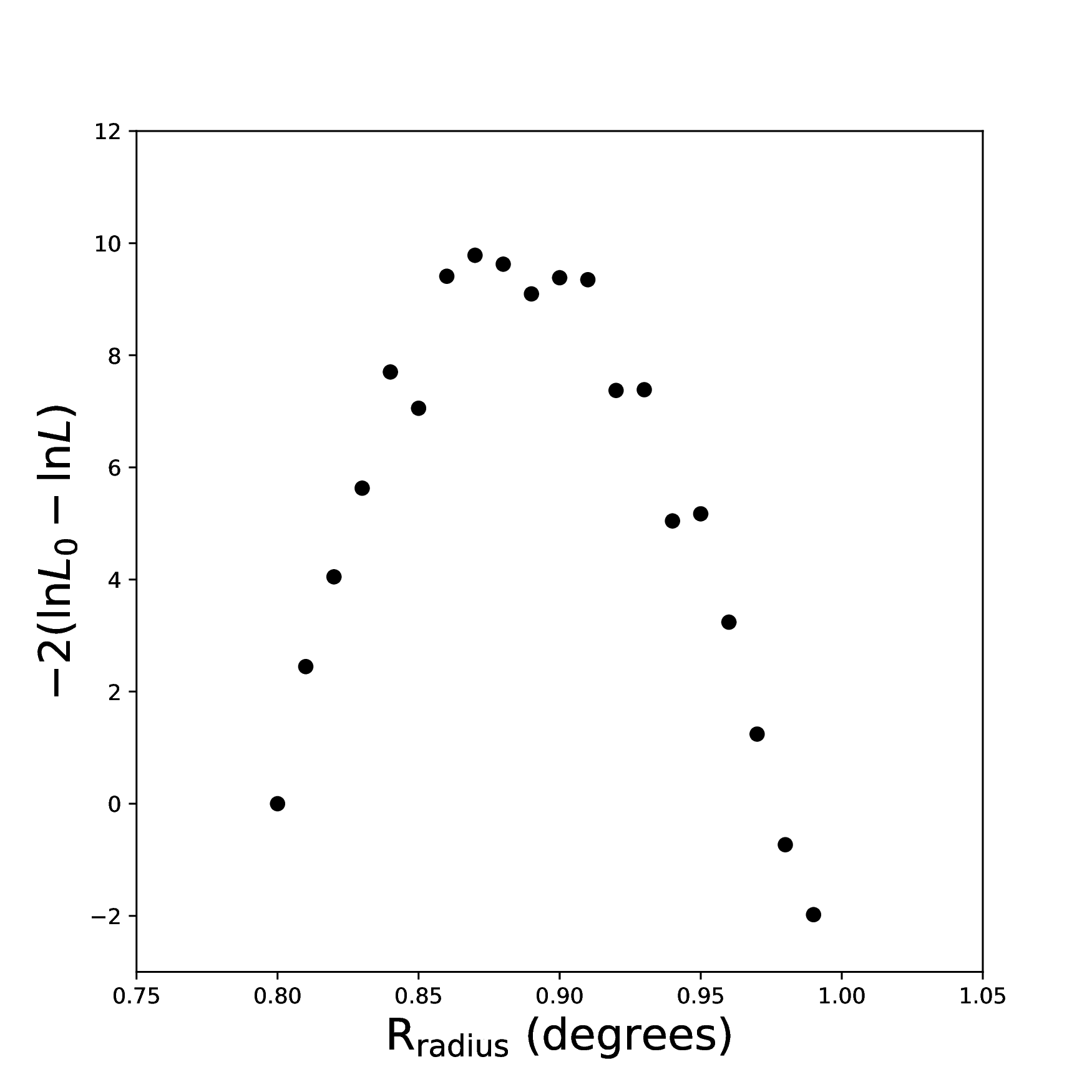}
        \caption {Significance of the uniform ellipse template A (left) and B (right) with various sizes ($L$) relative to the corresponding spatial template with the minimum size ($L_{0}$). The likelihood ratios are peaked at a semimajor axis of $\sim 0.49\deg$ and $\sim 0.87\deg$, respectively. See Sect.~\ref{sec:spatial_analy} for details.
}
\label{fig:likeratio}
\end{figure*}

To investigate the morphology and extension of the GeV \gray\ emission, we use the spatial template which is constitute of A, B, and HESS J1837-069, marked with the geometrics in Fig.~\ref{fig:spatialT}.
We note that HESS J1837-069 may be larger than the previously reported size, thus we use a uniform disk with a radius of $0.23\deg$ \citep{Aharonian06}.
For the region A, we produced several uniform elliptical disks centered at the position of the count peak value on the residual map with various semimajor axes from $0.4\deg$ to $0.6\deg$ in steps of $0.01\deg$ and a fixed semiminor axis of 0.25$\deg$.
The elliptical region B is centered at RSGC 1 with various semimajor axes from $0.8\deg$ to $1.0\deg$ in steps of $0.01\deg$ and a fixed semiminor axis of $0.5\deg$.
All the components are assumed to be a power-law spectral shape.
We first fixed the size of region B with a semimajor and semiminor axes of $0.9\deg$ and $0.5\deg$, and free the semimajor axis of region A. 
We performed the binned likelihood analysis on the events between 1 - 500 GeV energy band and obtained the best semimajor axis of region A.
Then we fixed the size of region A with the best size derived above and free the semimajor axis of region B.
We defined the uniform elliptical disk with various sizes as the alternative hypothesis ($L$), and the corresponding spatial disk with the minimum size as the null hypothesis ($L_{0}$), then compared the overall maximum likelihood of the $L$ with that of the $L_{0}$. Here we defined the significance of the alternative hypothesis model $-2({\rm ln}L_{0}-{\rm ln}L)$ following the paper in \cite{Lande12}.
As shown in the left and right panels of Fig.~\ref{fig:likeratio}, the likelihood ratio  peaks at the semimajor axis of $0.49^{\circ} \pm 0.02^{\circ}$ for the region A and $0.87^{\circ} \pm 0.04^{\circ}$ for the region B.   

\subsection{Spectral analysis}
\label{sec:spectral_analy}
The results from spatial analysis show that the semimajor axes of $\sim 0.49\deg$ for the region A and $\sim 0.87\deg$ for the region B are more preferred over the other elliptical sizes.
Thus we use the best spatial template as the spatial model of the extended GeV emission to extract the spectral information above 1 GeV energy band.

The obtained spectral parameters for the HESS J1837-069 are photon index of $1.83 \pm 0.08$, and energy flux of $(3.6 \pm 0.4) \times 10^{-11} \rm erg\ cm^{-2}\ s^{-1}$, corresponds to \gray\ luminosity of $\sim 1.9 \times 10^{35} (D/6.6 \ \rm kpc)^2 \ \rm erg\ s^{-1}$.
For the region A, the derived photon index is $2.05 \pm 0.02$ and the energy flux is $(1.6 \pm 0.1) \times 10^{-10} \rm erg\ cm^{-2}\ s^{-1}$, roughly equal to $\sim 2.4 \times 10^{35} (D/3.5 \ \rm kpc)^2 \ \rm erg\ s^{-1}$.
The photon index of region B,  $2.00 \pm 0.02$, is harder than that in MAGIC J1837-073, $\sim$ 2.3 above 300 MeV, \citep{MAGIC19}, and the energy flux is estimated to be $(2.4 \pm 0.1) \times 10^{-10} \rm erg\ cm^{-2}\ s^{-1}$ corresponding to \gray\ luminosity of $\sim 1.26 \times 10^{36} (D/6.6 \ \rm kpc)^2 \ \rm erg\ s^{-1}$.

%----------------------------------------------------- FIGURE 2

%----------------------------------------------------- FIGURE 3

%----------------------------------------------------- FIGURE 4
\begin{figure}
\centering
\includegraphics[scale=0.4]{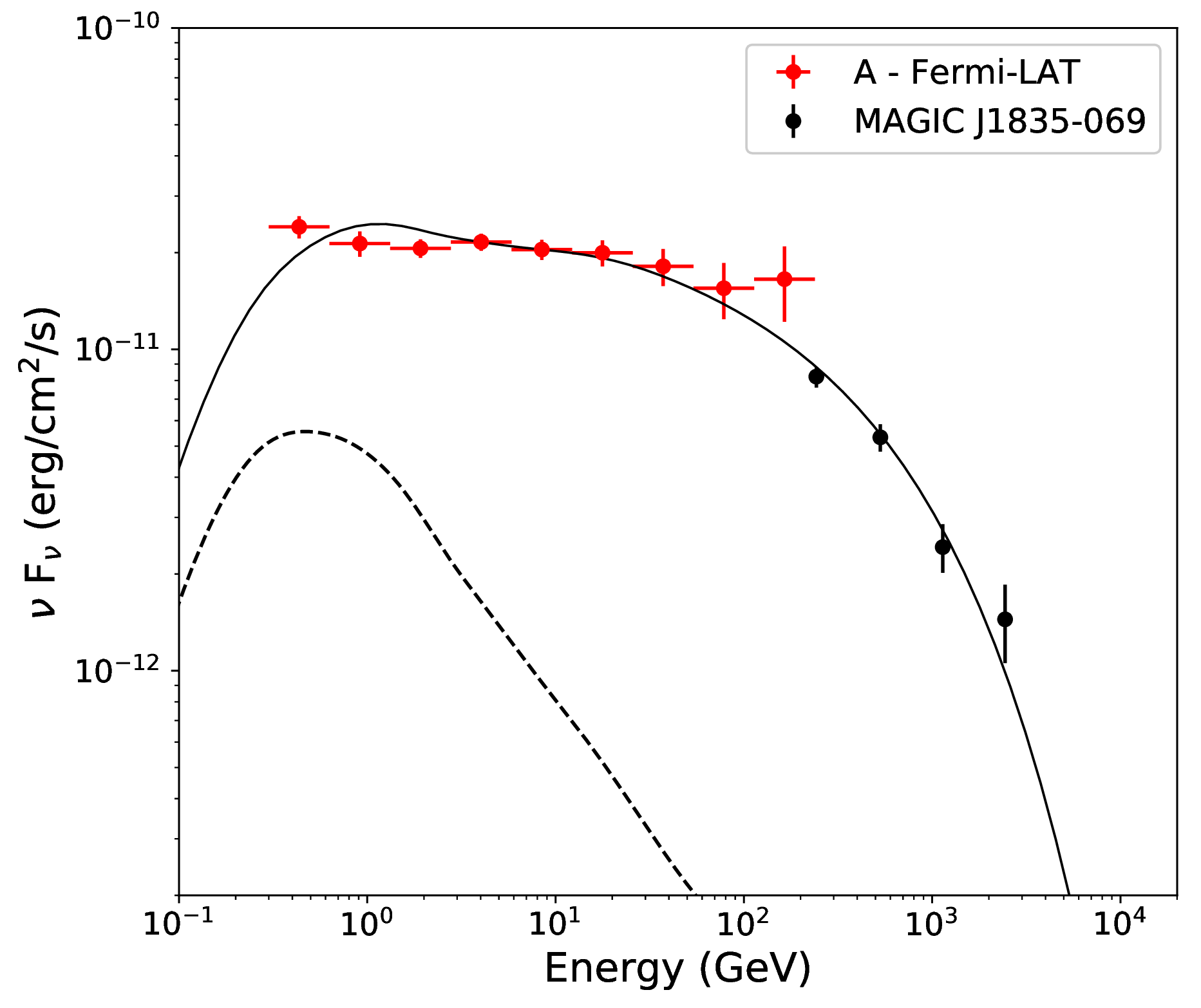}
%\caption {SEDs of GeV emission in the region A for a uniform elliptical spatial model with a semimajor and semiminor axes of $0.49\deg$ and $0.25\deg$. The data in the MAGIC J1835-069 region (black) are taken from \cite{MAGIC19}. The dashed line represents the predicted \gray\ emissions assuming the CR density in this region are the same as those measured locally by AMS-02 \protect \citep{ams02}.  See the context in Sect.~\ref{sec:Gas} for details.}
\caption {SEDs of GeV emission in the region A for a uniform elliptical spatial model with a semimajor and semiminor axes of $0.49\deg$ and $0.25\deg$. The data in the MAGIC J1835-069 region (black) are taken from \protect \cite{MAGIC19}. The dashed line represents the predicted \gray\ emissions assuming the CR density in this region are the same as those measured locally by AMS-02 \protect \citep{ams02}.  See the context in Sect.~\ref{sec:Gas} for details.}
\label{fig:A}
\end{figure}

%----------------------------------------------------- FIGURE 5
\begin{figure}
%\centering
\includegraphics[scale=0.4]{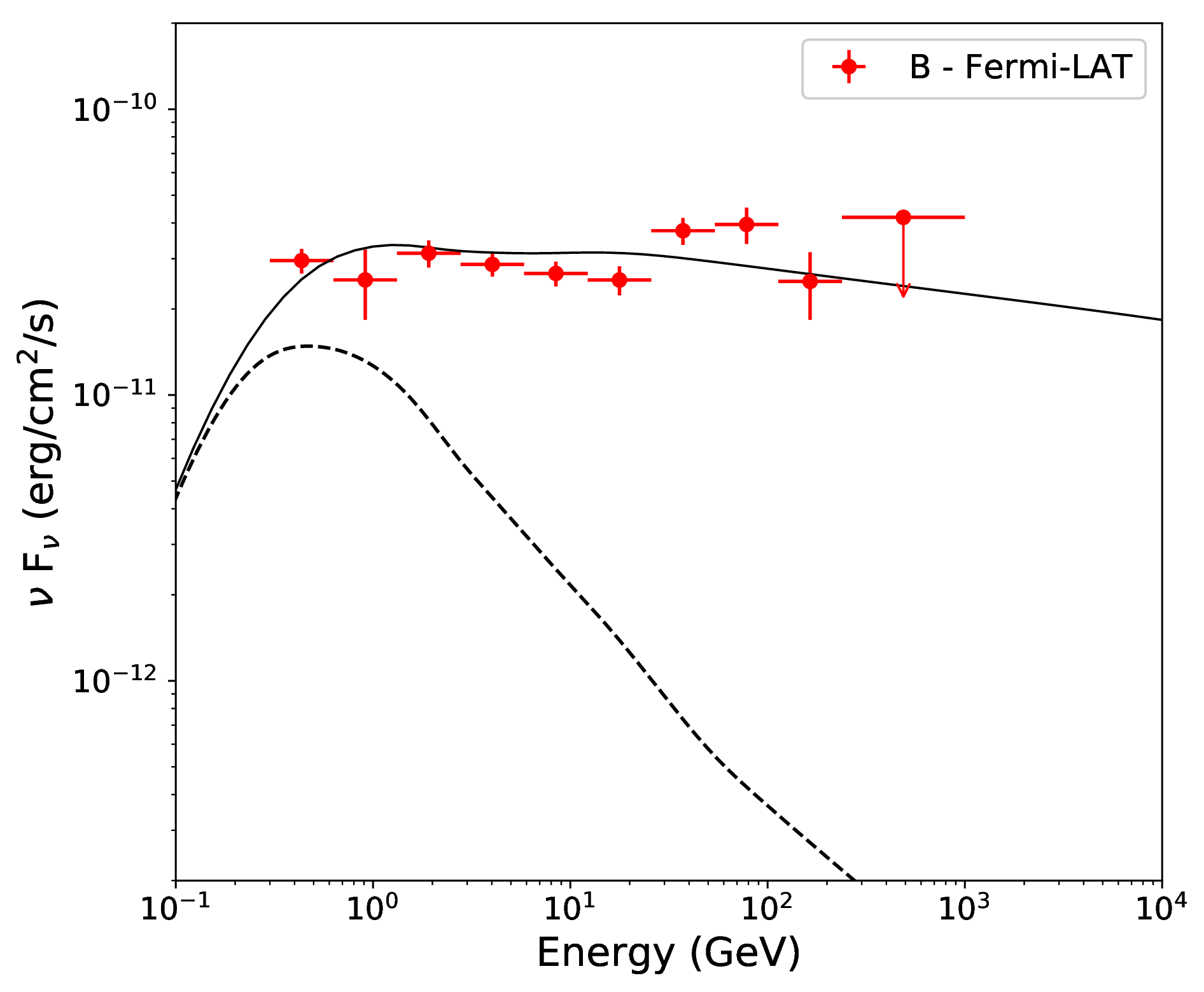}
\caption {
SED of \gray\ emission around RSGC 1 (B region) for a uniform elliptical spatial model with a semimajor and semiminor axes of $0.87\deg$ and $0.5\deg$.
The dashed line represents the predicted \gray\ emissions assuming the CR density in this region are the same as those measured locally by AMS-02 \citep{ams02}.}
\label{fig:B}
\end{figure}

%----------------------------------------------------- FIGURE 6
\begin{figure}
%\centering
\includegraphics[scale=0.4]{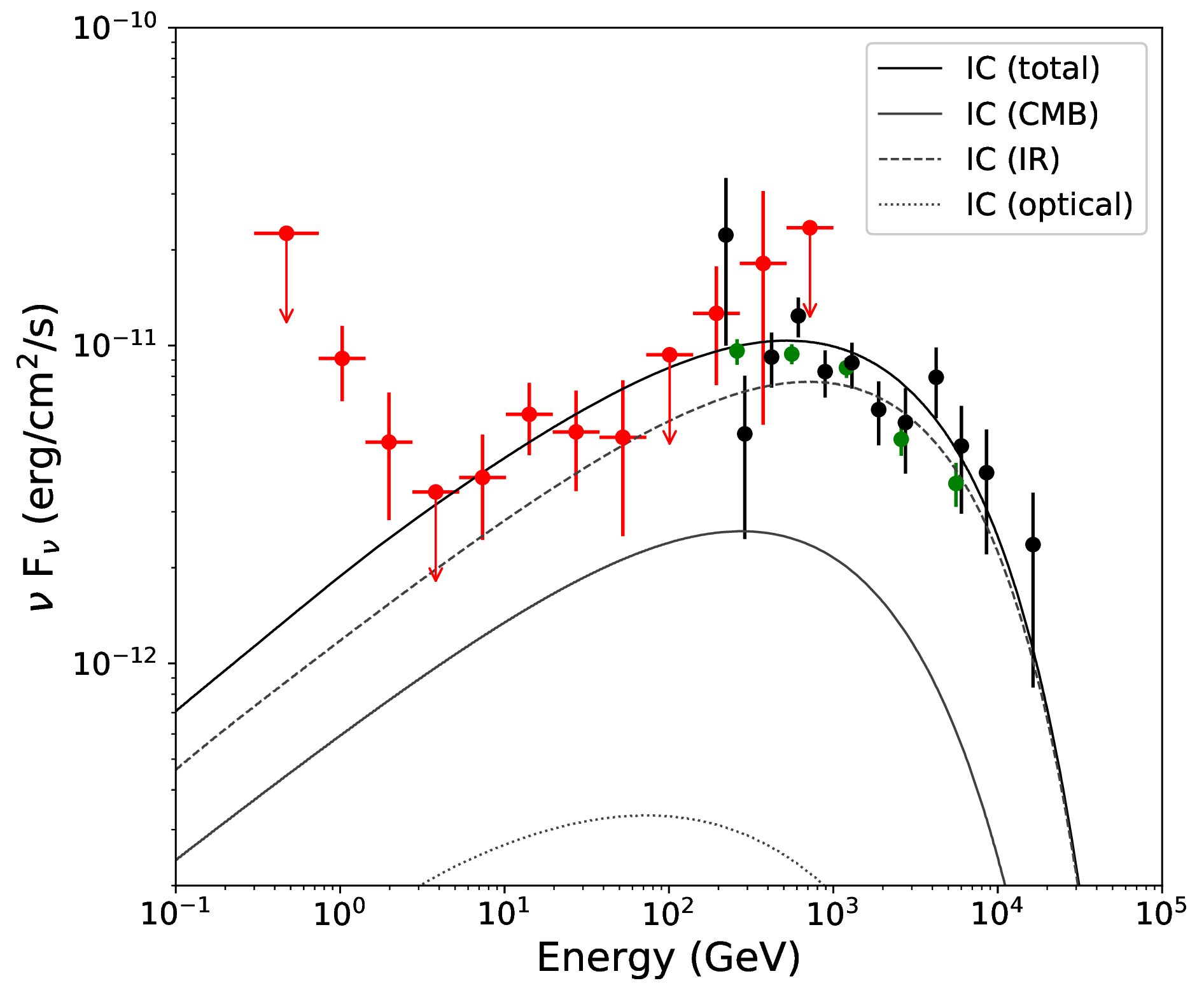}
\caption {SED of the HESS J1837-069 region measured by the \fermi\ (red points with statistical errors of $1\sigma$). 
For the energy bins with $\rm TS < 4$, the upper limits are calculated within a 3$\sigma$ confidence level.
The black and green data of HESS J1837-069 are taken from \protect \cite{Aharonian06} and \protect \cite{MAGIC19}, respectively.
}
\label{fig:HESS069}
\end{figure}

%----------------------------------------------------- FIGURE 7
\begin{figure*}
%\centering
\includegraphics[scale=0.27]{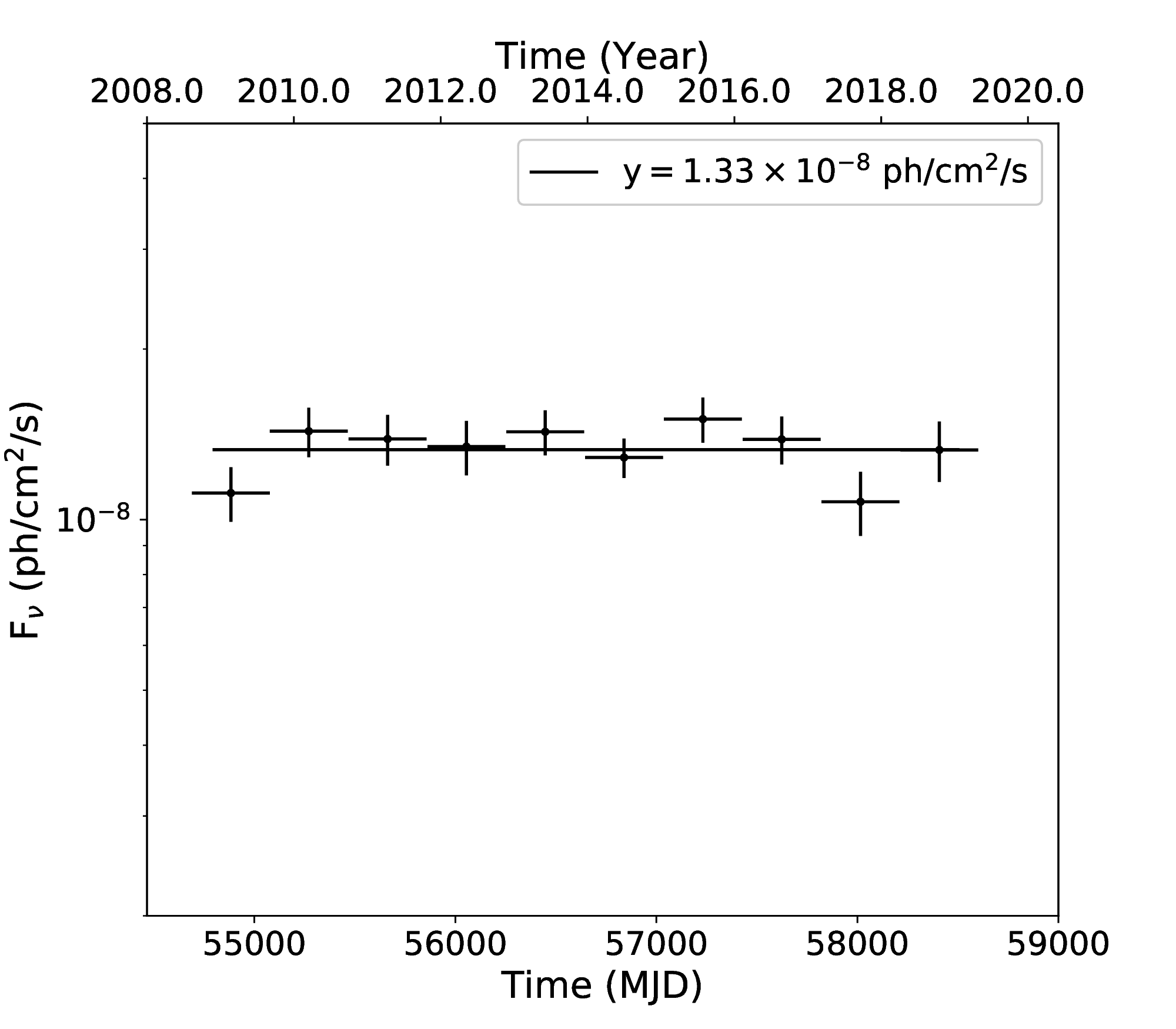}
\includegraphics[scale=0.27]{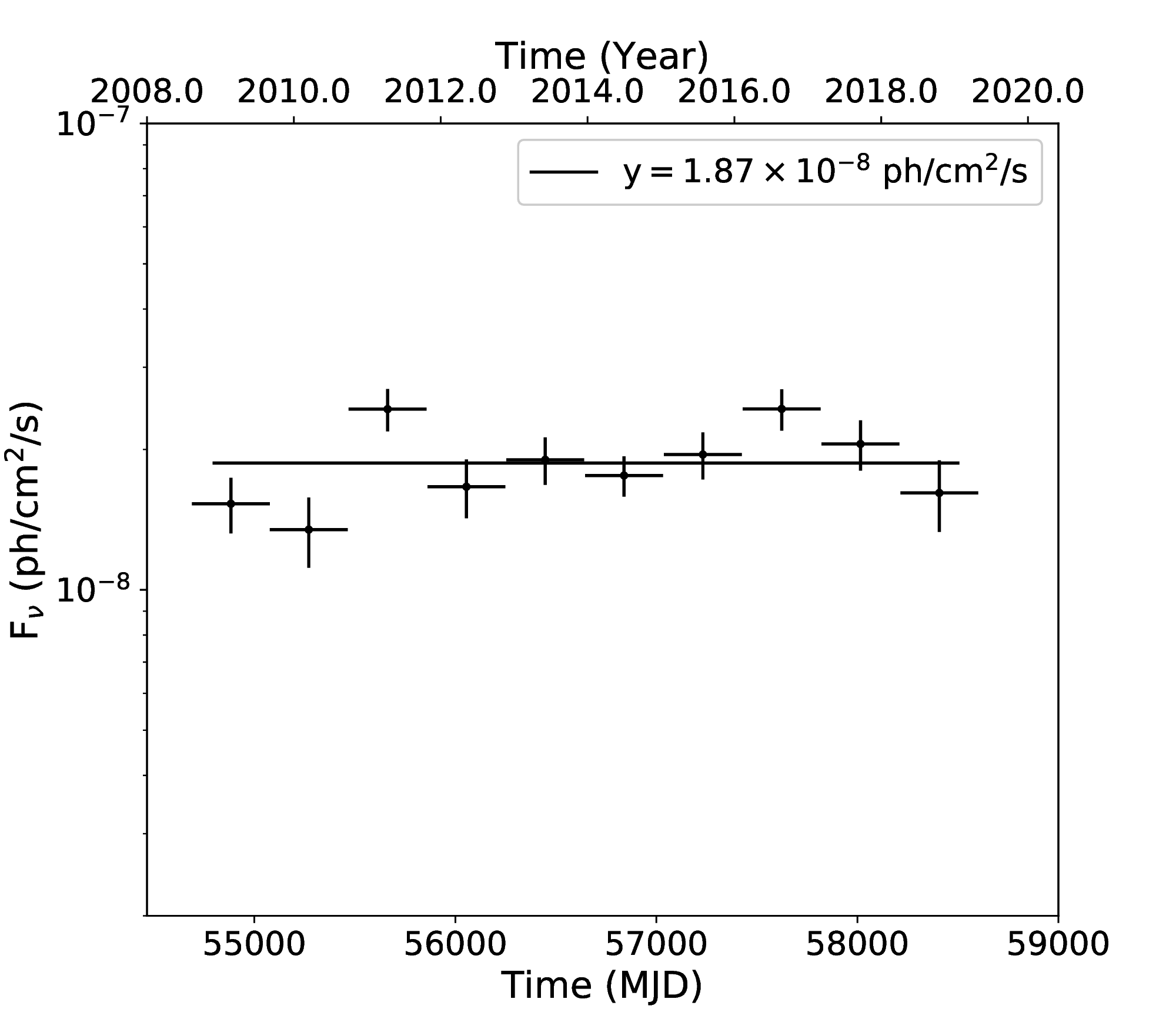}
\includegraphics[scale=0.27]{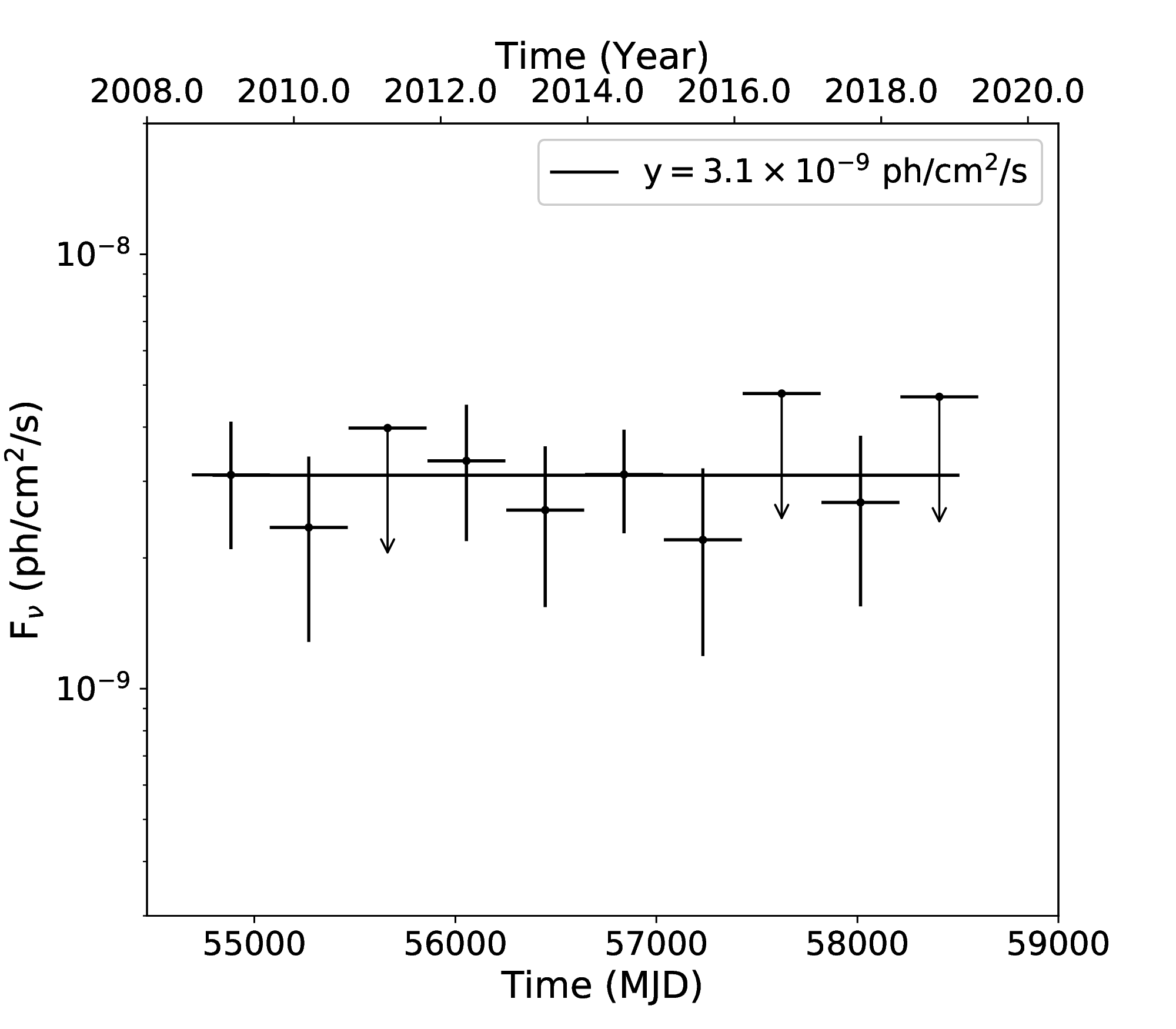}
\caption {
Light curves of the \gray\ emission from region A, B, and HESS J1837-069 from August 4, 2008 (MJD 54683) until August 03, 2019 (MJD 58698).
The horizontal lines are the best fitting lines correspondingly.
}
\label{fig:lc}
\end{figure*}

To study the origin of the GeV emission and the underlying particle spectrum that give rise to the observed spectrum of photons, we derive the spectral energy distributions (SEDs) via the maximum likelihood estimation in logarithmically spaced bins for the \gray\ emission above 300 MeV.
Fig.~\ref{fig:A} - Fig.~\ref{fig:HESS069} show the resulting SEDs of the extraction region A, B, and HESS J1837-069.
Fig.~\ref{fig:A} shows the SED of the region A (red) measured by \fermi\ combined with the SED of the MAGIC source (black) \citep{MAGIC19}.
The smoothly connection with the SED of MAGIC J1835-069 above 200 GeV implies a possible common origin, which is in good agreement with that of \cite{MAGIC19}.
Fig.~\ref{fig:B} shows that the SED of the region B  has a very hard spectrum above $\sim 300\ \rm MeV$.
Fig.~\ref{fig:HESS069} shows the SED of HESS J1837-069 obtained with \fermi\ (red) combined with the SEDs of the same region measured by HESS (black) and MAGIC (green) \citep{Aharonian06,MAGIC19}.
 If the significance of the energy bin is less than $2\sigma$, we only free the normalization parameter obtained from the broadband fitting and calculate the upper limit within $3\sigma$ confidence level.

To test the possible variabilities of the three regions, we produce the light curves by binning the whole data set used in the analysis into 10 equidistant time bins, and derive the \gray\ spectrum above 1 GeV in each of these bins.
The results are shown in Fig.\ref{fig:lc}.
For the time bins, which are detected with a significance of less than $2\sigma$, we calculate the upper limits within $1\sigma$ confidence level.
By fitting the light curves with a horizontal line, the derived reduced chi-squared $\rm \chi^2/dof = 0.96, 2.08$ and $1.71$, which are basically consistent with the static emission.

\section{Gas content around RSGC 1}
\label{sec:Gas}
We study three different gas phases, i.e., the neutral atomic hydrogen (\ion{H}{i}), the molecular hydrogen (H$_{2}$), and the ionized hydrogen (\ion{H}{ii}), in the vicinity of the RSGC 1 region.

We derive the \ion{H}{i} column density $N_{\ion{H}{i}}$ from the data-cube of Galactic \ion{H}{i} 4$\pi$ survey (HI4PI) \citep{HI4PI16} using the expression, 
\begin{equation}
N_{\ion{H}{i}}=-1.83 \times 10^{18}T_{\rm s}\int \mathrm{d}v\ {\rm ln} \left(1-\frac{T_{\rm B}}{T_{\rm s}-T_{\rm bg}}\right),
\end{equation}
where $T_{\rm bg} \approx 2.66\ \rm K$ is the brightness temperature of the cosmic microwave background (CMB) radiation at 21-cm, and $T_{\rm B}$ is the measured brightness temperature in 21-cm surveys.
In the case of $T_{\rm B} > T_{\rm s} - 5\ \rm K$, we truncate $T_{\rm B}$ to $T_{\rm s} - 5\ \rm K$, and a uniform spin temperature $T_{\rm s}$ is chosen to be $150 \ \rm K$.

To trace the H$_{2}$, we use the Carbon Monoxide (CO) composite survey \citep{Dame01} and the standard assumption of a linear relationship between the velocity-integrated brightness temperature of CO 2.6-mm line, $W_{\rm CO}$, and the column density of molecular hydrogen, $N(\rm H_{2}$), i.e., $N({\rm H_{2}}) = X_{\rm CO} \times W_{\rm CO}$ \citep{Lebrun83}.
The conversion factor $X_{\rm CO} = \rm 2.0 \times 10^{20}\ cm^{-2}\ K^{-1}\ km^{-1}\ s$ has been adopted \citep{Bolatto13,Dame01}.

\begin{figure*}
%\centering
\includegraphics[scale=0.35]{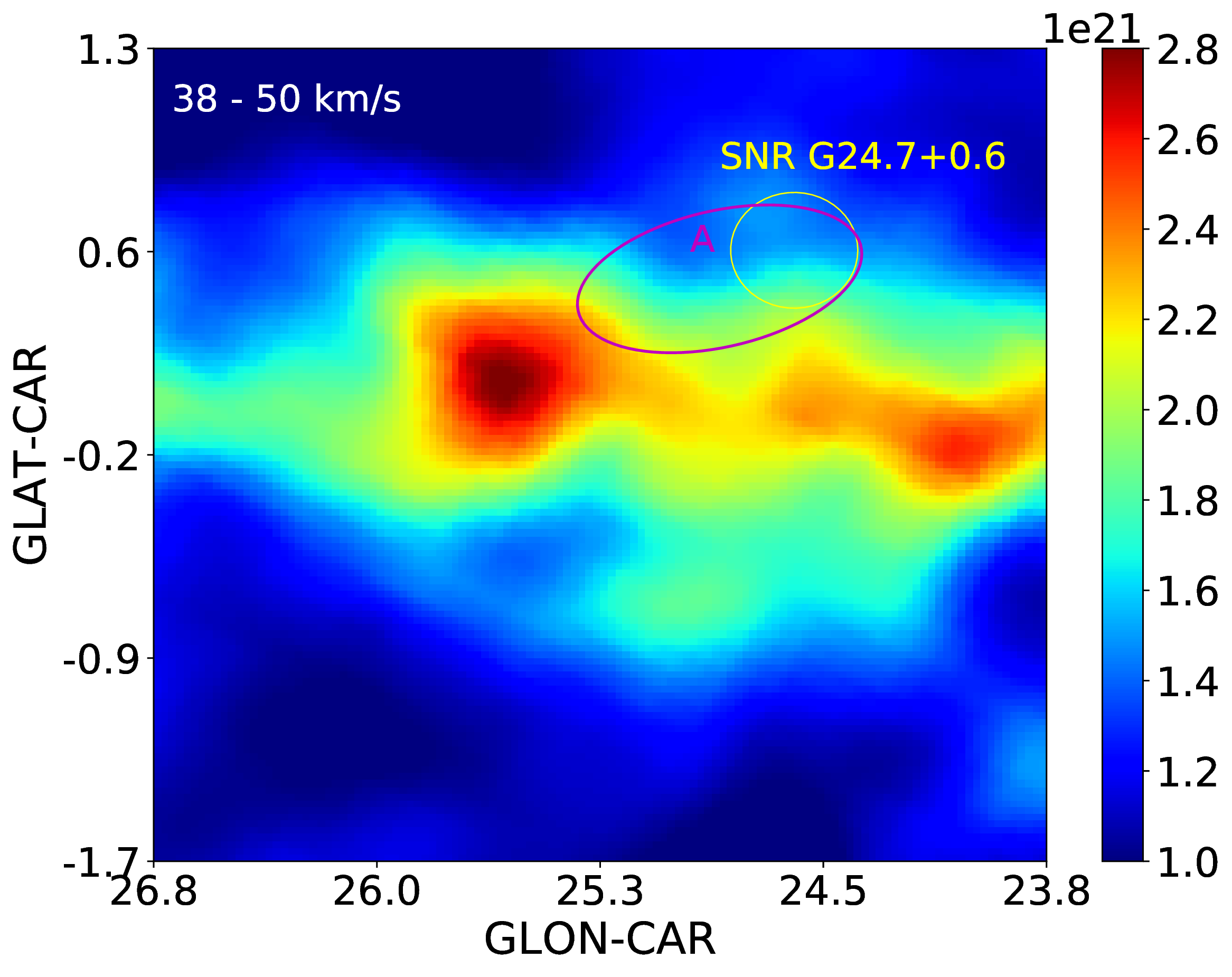}
\includegraphics[scale=0.35]{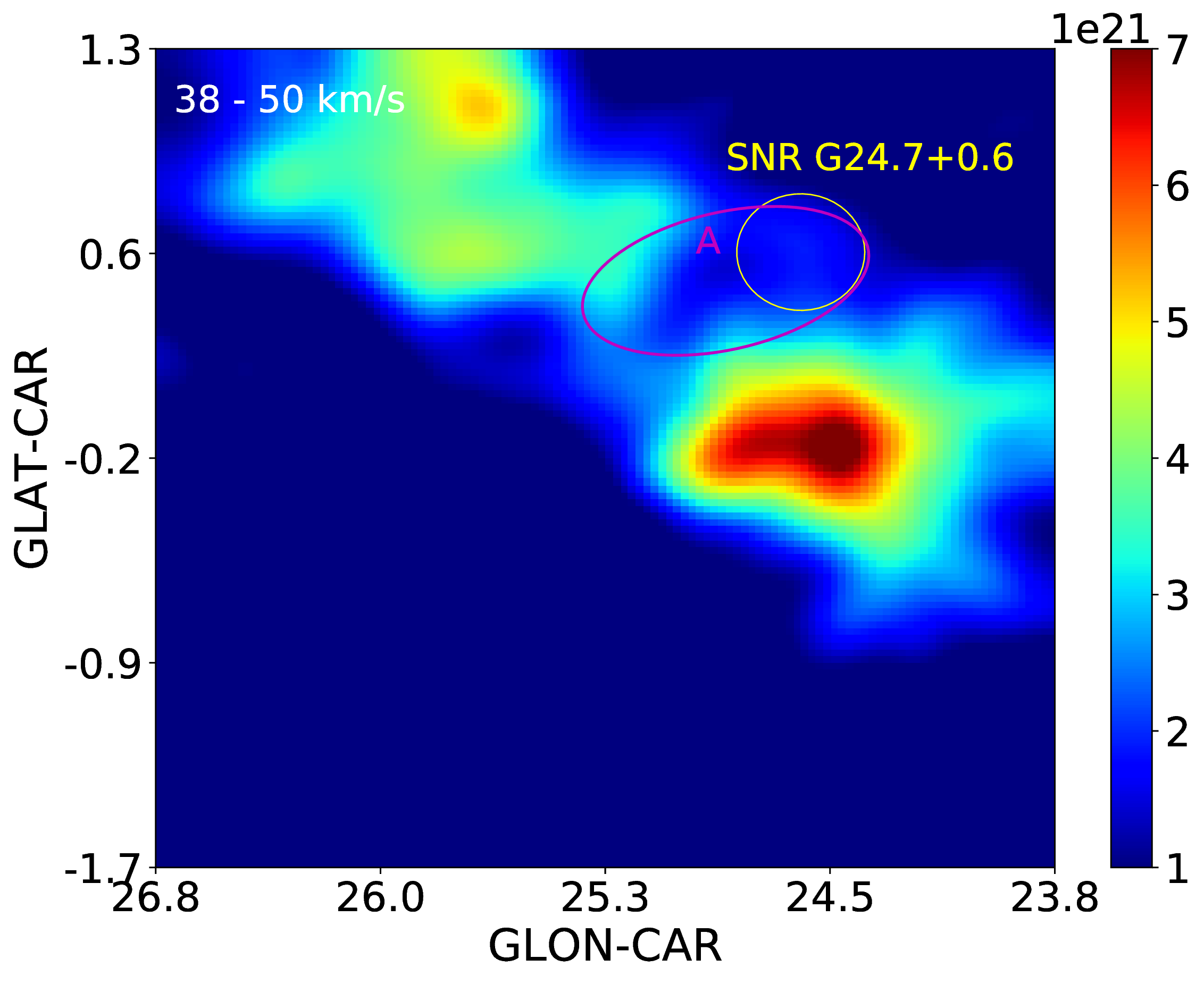}
\caption {
Left shows the map of \ion{H}{i} column density derived from 21-cm all-sky survey. Right shows the H$_{2}$ column density derived from the CO data.
Integrated the gases within the velocity range of $38 - 50\ \rm km~s^{-1}$. 
The red ellipse represents the region A and the yellow circle marks the GeV \gray\ emission of the SNR G24.7+0.6 listed in 4FGL.
See the context in Sect.~\ref{sec:Gas} for details.
}
\label{fig:gasdis1}
\end{figure*}

\begin{figure*}
%\centering
\includegraphics[scale=0.35]{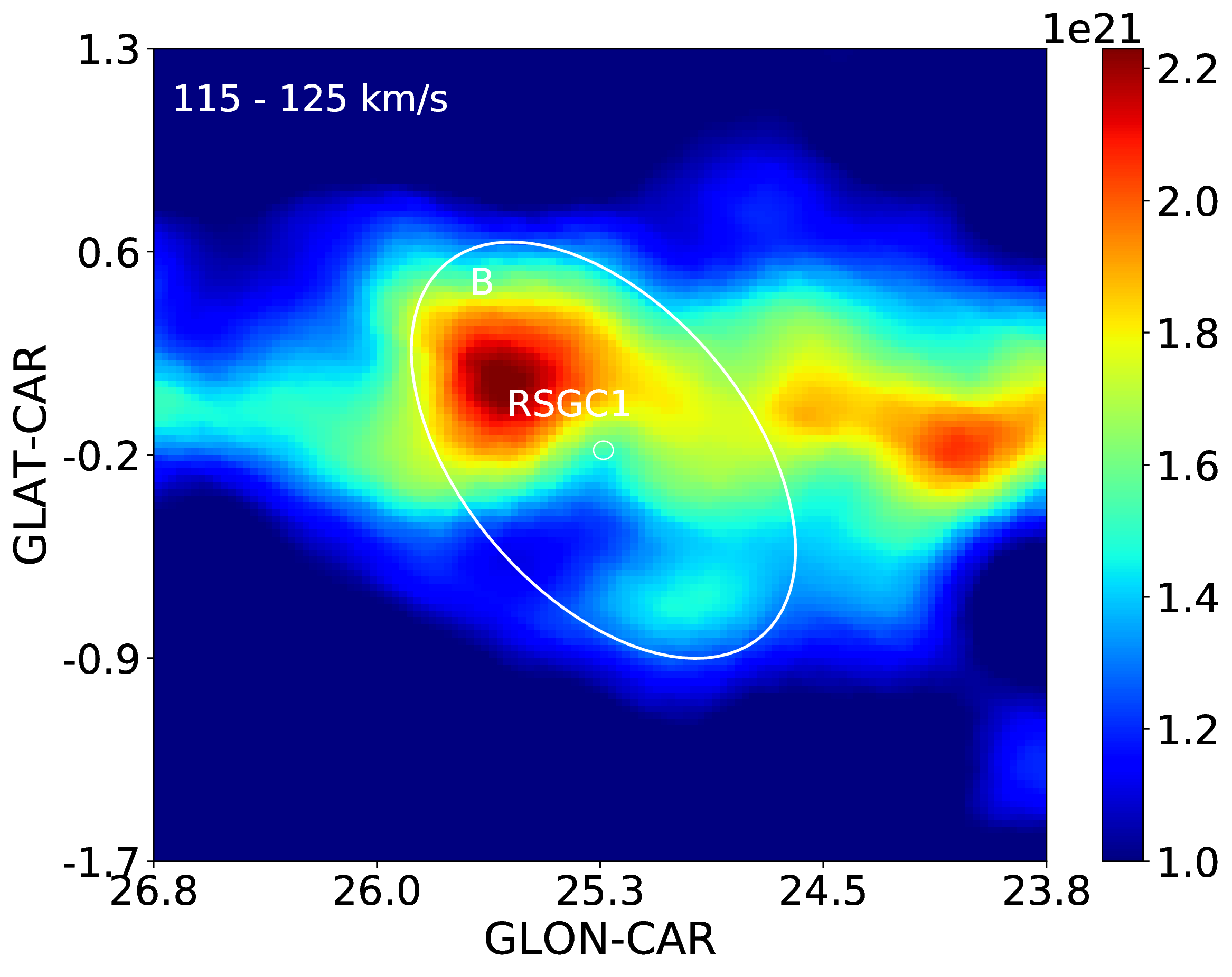}
\includegraphics[scale=0.35]{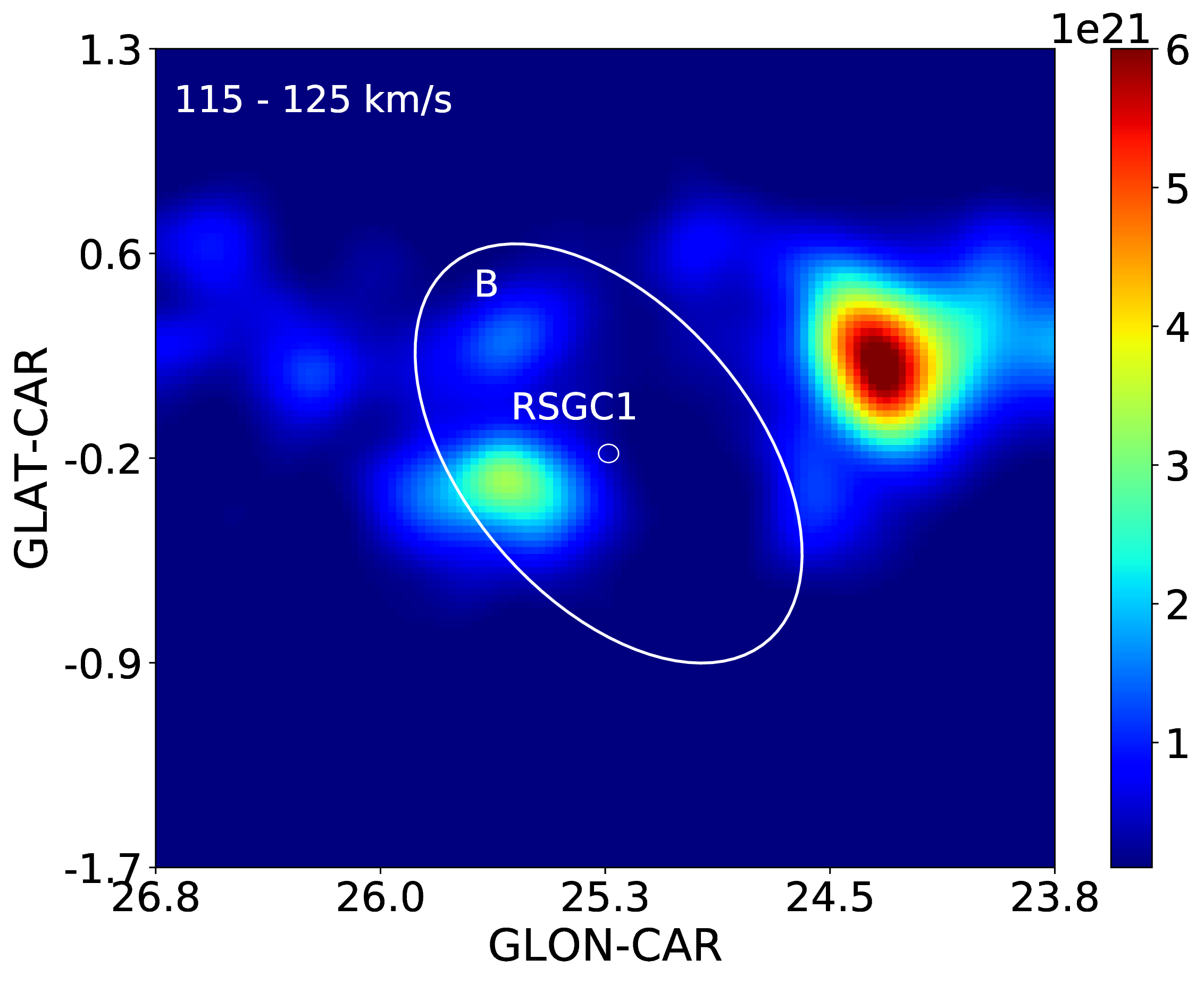}
\caption {
Same as Fig.~\ref{fig:gasdis1}, but integrated the velocity interval from $115 - 125\ \rm km~s^{-1}$. The white circle is the approximate extent of the massive star cluster RSGC 1 and the white ellipse represents the GeV emission from the region B.
}
\label{fig:gasdis2}
\end{figure*}

The distance to the SNR G24.7+0.6 is 3.5 kpc, and the radial velocity of the gas in its surroundings is in the range of $\nu_{\rm LSR} \sim 38 - 50\ \rm km~s^{-1}$ \citep{Petriella08,Petriella10}.
Thus we investigate the gas distributions around the region A based on the above values, and the derived column density maps of \ion{H}{i} and H$_{2}$ are shown in  the left and right panels of Fig.~\ref{fig:gasdis1}, respectively.
To study the gas distributions within the region B, we adopt the measured radial velocities of the stars in the cluster RSGC 1, in the range of $115\ \rm km~s^{-1}$ to $125\ \rm km~s^{-1}$, and a kinematic distance of $6.60 \pm 0.89~\rm kpc$ given in \cite{Davies08}.
The obtained gas distributions of \ion{H}{i} and H$_{2}$ are shown in the left and right panels of Fig.~\ref{fig:gasdis2}, respectively.

\begin{figure}
%\centering
\includegraphics[scale=0.35]{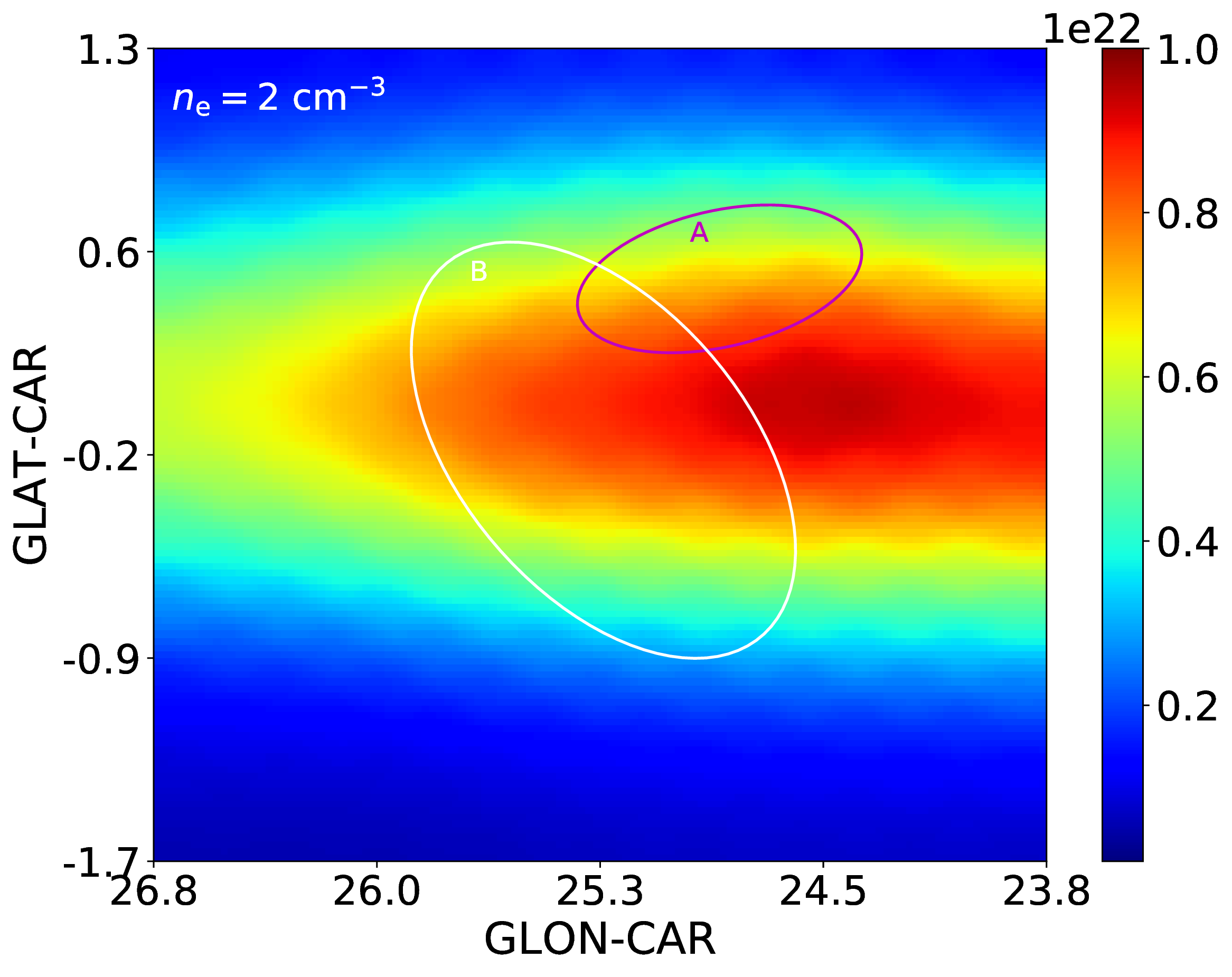} 
\caption {
\ion{H}{ii} column density derived from the \planck\ free-free map assuming the effective density of electrons $n_{\rm e}=2~\rm cm^{-3}$.
The red and white ellipses represent the region A and B.
For details, see the context in Sect.~\ref{sec:Gas}.
}
\label{fig:gasdis3}
\end{figure}

We select the free-free emission map derived from the joint analysis of {\it Planck}, {\it WMAP}, and 408 MHz observations \citep{Planck16} to derive the map of \ion{H}{ii} column density.
First we convert the emission measure into free-free intensity ($I_{\nu}$) by using the conversion factor at 353-GHz in Table 1 of \citet{Finkbeiner03}.
Then we utilize the Equation (5) in \citet{Sodroski97},
\begin{equation}
\begin{aligned}
N_{\ion{H}{ii}} = &1.2 \times 10^{15}\ {\rm cm^{-2}} \left(\frac{T_{\rm e}}{1\ \rm K}\right)^{0.35} \left(\frac{\nu}{1\ \rm GHz}\right)^{0.1}\left(\frac{n_{\rm e}}{1\ \rm cm^{-3}}\right)^{-1} \\
&\times \frac{I_{\nu}}{1\ \rm Jy\ sr^{-1}},
\end{aligned}
\end{equation}
to convert the free-free intensity into column density in each pixel, with a frequency at $\nu = \rm 353\ GHz$, and an electron temperature of $T_{e} =\rm 8000\ K$. 
This equation also shows that the \ion{H}{ii} column density is inverse proportional to the effective density of electrons $n_{\rm e}$, thus we choose $2\ \rm cm^{-3}$ \citep{Sodroski97} to calculate the upper limit of the \ion{H}{ii} column density, which is shown in the Fig.~\ref{fig:gasdis3}.

%
%\cite{Rahman10} estimated the distance to the "G25 bubble" \citep{Katsuta17} is 6.1 kpc, which is probably an underestimate because they are based on the median velocity ($\sim 100~\rm km/s$) of the \ion{H}{ii} regions.
%Thus we selected a greater distance of 7.7 kpc and a radial velocity range of $\nu_{\rm LSR} \sim 95 - 125 \rm km/s$ for the G25 region \citep{Katsuta17}.

%
We calculate the total mass within the cloud in each pixel from the expression
\begin{equation}
M_{\rm H} = m_{\rm H} N_{\rm H} A_{\rm angular} d^{2}
\end{equation}
where $m_{\rm H}$ is the mass of the hydrogen atom, $N_{\rm H} = N_{\ion{H}{ii}} + 2N_{\rm H_{2}} + N_{\ion{H}{i}}$ is the total column density of the hydrogen atom in each pixel.
$A_{\rm angular}$ refers to the angular area, and $d$ is the distance of the target cloud (region A or B).
We calculate the mass and number of the hydrogen atom in each pixel, and then estimate the mass and number summations within the region A and B (see Fig.~\ref{fig:spatialT}).
The total mass is estimated to be $\sim 2.66 \times 10^{5}~\msun$ for the region A and $\sim 2.53 \times 10^{6}~\msun$ for the region B.
Assuming ellipsoidal geometries of the GeV \gray\ emission within both the region A and B, with the corresponding sizes of ($0.49\deg,\ 0.25\deg$) and ($0.87\deg,\ 0.50\deg$), the volumes of the two ellipsoids are estimated as $V = \frac{4\pi}{3}r_{\rm 1}r_{\rm 2}^{2}$.
Here $r_{\rm 1,2} = d\times\theta_{\rm 1,2}(\rm rad)$ are the semimajor and semiminor axes of the assumed ellipses, $d$ is the distance to the objective region.
Then the average gas number densities over the ellipsoidal volumes of region A and B are calculated to be $\rm \sim 371\ cm^{-3}$ and $\rm \sim 73\ cm^{-3}$, respectively.
Table \ref{tab2} shows the gas total masses and number densities within the region A and region B.

%----------------------------------------------------- FIGURE 8

%----------------------------------------------------- FIGURE 9

%----------------------------------------------------- FIGURE 10
%

%----------------------------------------------------- TABLE 2
\begin{table}
%\centering
\caption{Gas total masses and number densities within the region A and region B. See Sect.~\ref{sec:Gas} for details.}
\begin{tabular}{lcc}
\hline
\hline
Region & Mass ($10^{5}\msun$) & Number density ($\rm cm^{-3}$) \\
\hline
A& 2.66&371\\ %\cline{1-1}
B&25.3&73\\ %\cline{1-1}
\hline
\hline
\end{tabular}\\
\label{tab2}
\end{table}

\section{The origin of the \gray\ emission}
\label{sec:cr}
In the following, we utilize {\it Naima}\footnote{\url{http://naima.readthedocs.org/en/latest/index.html}}. \citep{naima} to fit the SEDs. Naima is a numerical package that includes a set of nonthermal radiative models and a spectral fitting procedure, which allow us to implement different functions and perform Markov chain Monte Carlo \citep[MCMC;][]{Foreman13} fitting of nonthermal radiative processes to the data.

\subsection{HESS J1837-069}
\cite{Fujita14} proposed that the TeV emission of HESS J1837-069 is most likely from the PWN hidden in this region.
\citet{Gotthelf08} found two powerful PWNs in this region, AX J1837.3-0652 and AX J1838.0-0655, both of which were powered by the pulsar with the spin-down luminosity of more than  $4\times 10^{36}~\rm erg/s$.  The \fermi\ data extend the spectrum to lower energy.
To phenomenologically interpret the SED of HESS J1837-069, we adopt a leptonic model where the \grays\ are from the Inverse Compton (IC) scattering.
During the fitting we exclude the first three data points, which maybe originate from another  radiation mechanism.
We assume that the electrons have a cutoff power-law function,
\begin{equation}
        N(E) = A~E^{-\alpha}~{\rm exp}\left(-\frac{E}{E_{\rm cutoff}}\right),
\label{equ:ecpl}
\end{equation}
treating $A$, $\alpha$, and $E_{\rm cutoff}$ as free parameters for the fitting.
The target photon fields for relativistic electrons to scatter around the position of HESS J1837-069 include the CMB, infrared, and starlight photons adopted from the local interstellar radiation field calculated in \citet{Popescu17}.
The leptonic model can explain the observed \grays\ shown in Fig.~\ref{fig:HESS069} with a maximum log-likelihood value (MLL) of -9.4.
The obtained total energy of the electrons (>10 GeV) is $\rm W_{e} = (4.5 \pm 0.9) \times 10^{48}~\rm erg$, with the index $\alpha = 2.21 \pm 0.09$, and $E_{\rm cutoff} =\rm 10 \pm 2\ TeV$.
From the fitted results we found a cutoff power-law type spectrum of electrons under the assumption that the GeV--TeV \gray\ emission is due to the IC scattering, which is compatible with the one-zone PWN emission model. Taking into account the age of PSR J1838-0655 of about 23\,kyr and the cutoff energy of about 10 TeV, the magnetic field should be close to $15\,\rm \mu G$. It is also possible the PWN AX J1837.3-0652 also contribute significantly to the \gray\ emission. In this case due to the fact that the pulsar powering AX J1837.3-0652 can be significantly older \citep{Gotthelf08} we expect a significant lower magnetic field in this region.

\subsection{Region A}
%\subsection{Modeling the spectral energy distribution}
%
In our analysis, the position of region A reveals a better spatial correlation with  MAGIC J1835-069 than FGES J1834.1-0706 as in \citet{MAGIC19}. The slight difference may come from the updated Fermi diffuse background. The SEDs are compatible with the results in \citet{MAGIC19}. Thus both spatial and spectral results are in consistency with the model proposed in \citet{MAGIC19} to explain the \gray\ emission from MAGIC J1835-069, that is, the CRs escaped from SNR G24.7+0.6 interacting with the surrounding gas. 

To estimate the distributions of the relativistic particles that are responsible for the GeV \gray\ emission in the region A, we consider a hadronic model, in which the high energy \grays\ produced in the pion-decay process following the proton-proton inelastic interactions.
We simply assume that the spectral distribution of the radiating protons has the same form as the equation~\ref{equ:ecpl}.
The average number densities of the target protons for the region A and B are assumed to be $\rm 371\ cm^{-3}$ and $\rm 73\ cm^{-3}$ derived from the gas distributions in Sect.~\ref{sec:Gas}.

In Fig.~\ref{fig:A} we present the best-fit results for region A. The derived total energy of the protons (>2 GeV) is $\rm W_{p} = (2.7 \pm 0.1) \times 10^{48}~\rm erg$, with the index $\alpha = 2.20 \pm 0.02$, and $E_{\rm cutoff} =\rm 7.4 \pm 1.7\ TeV$.   
The dashed line in Fig.~\ref{fig:A} represents the predicted \gray\ emissions derived assuming the CR density in region A is the same as the local measurement by AMS-02 \citep{ams02}. The results reveal a significant CR enhancement in this region, which is predicted near the CR acceleration site. 

We cannot formally rule out the possibility that \grays\ in region A are from one unknown PWN. But as estimated in \citet{MAGIC19}, a unrealistic high spin-down luminosity is required to explain the \gray\ emission, which makes such scenario quite unlikely.  

\subsection{Region B}
\citet{Katsuta17} also investigated the diffuse \gray\ emission in this region  and argued that it can be a clone of Cygnus cocoon, i.e.,  the CR cocoon/bubble formed by the young OB star clusters. They further found a OB association candidate G25.18+0.26  with the X-ray observations. In our updated analysis we further separated region A from region B due to the spatial coincidence of Region A with the TeV source MAGIC J1835-069. The resulted diffuse emission labeled as region B has a significantly different morphology from that in \cite{Katsuta17}. In our case the young cluster RSGC 1 centered in region B. Although we cannot rule out the possibility that G25.18+0.26 is the source of the CRs which produced the \gray\ emissions, we postulate here that RSGC 1 may be a more natural source to accelerate the CRs.   Indeed,  RSGC 1 is one of the most massive young star clusters in our Galaxy. It has an age of about 10~Myrs, which is older than the other known young star clusters which have already been detected in \grays.\ But it still contains more than 200 main-sequence massive stars (with a mass of more than $8~M_{\odot}$ ) \citep{froebrich13}.   The stellar wind of these massive stars may  provide enough power to accelerate the CRs. 
Since the \grays\ reveal no clear curvature we assume a single power-law spectrum for parent protons in deriving the CR content, i.e., 
  \begin{equation}
	N(E) = A~E^{-\alpha},
\label{equ:pl}
\end{equation}
the derived parameters are $\alpha = 2.15 \pm 0.03$, and the total energy is $\rm W_{p} = (9.6 \pm 0.6) \times 10^{49}~\rm erg$ for the protons above 2 GeV. 
%

%----------------------------------------------------- FIGURE 11
\begin{figure}
%\centering
\includegraphics[scale=0.4]{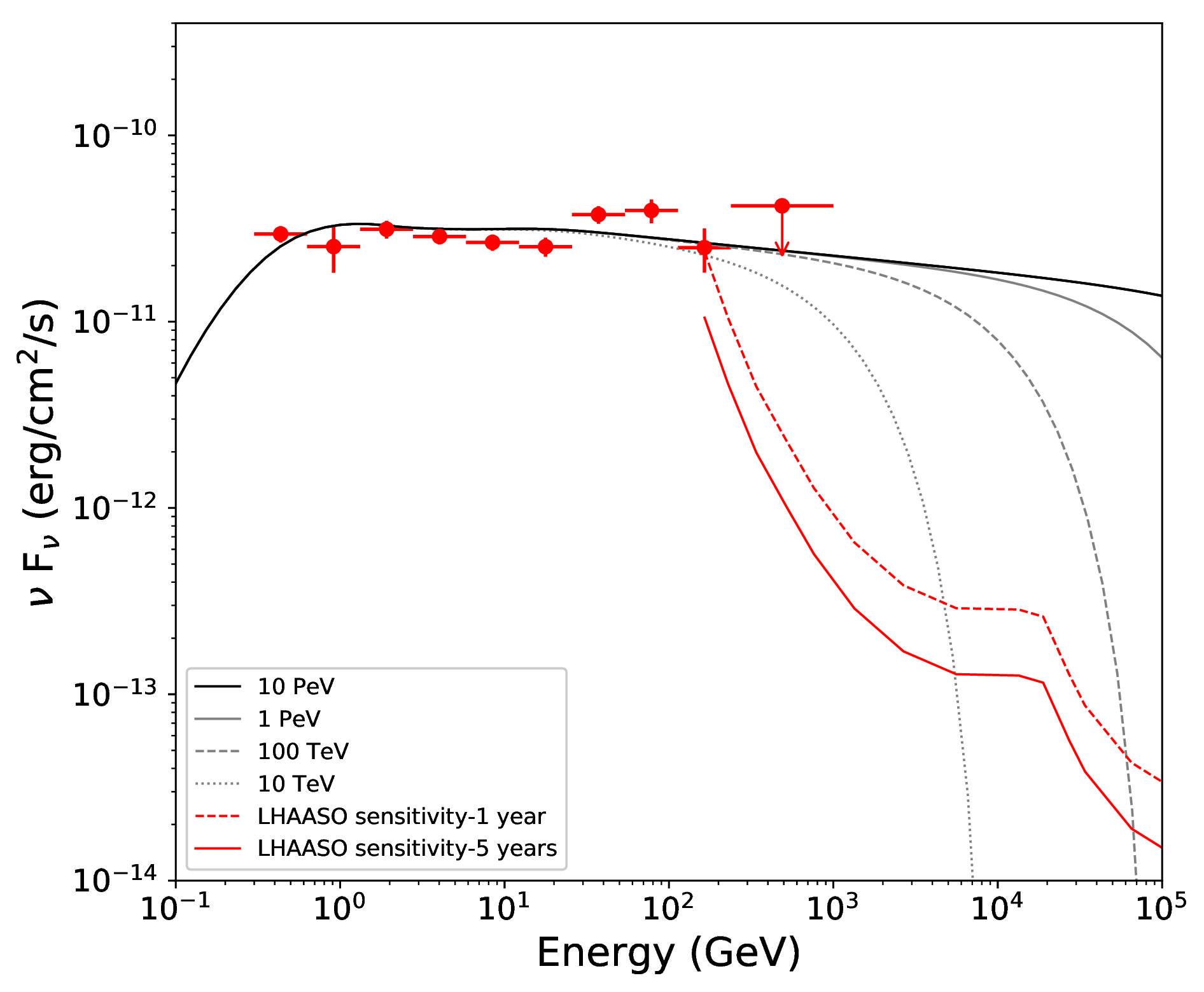} 
\caption {
The LHAASO detectability of region B. The red curves are LHAASO sensitivity of  the extended source with the same size of region B for 1 year (dashed) and 5 years (solid) observations. Also plotted is the fitted SEDs with different cutoff energy of parent protons.  }
\label{fig:lhaaso}
\end{figure}

\section{Discussion and conclusion}\label{sec:conc}
In this paper, we performed detailed analysis of Fermi LAT data on the vicinity of the young massive cluster RSGC 1. We found that the \gray\ emission in this region can be resolved into three components.  First we found GeV \gray\ emission coincide with the VHE \gray\ source HESS J1837-069. It has a photon index of $1.83\pm0.08$ at GeV band. Combining this with the HESS and MAGIC data, we argue that the \gray\ emission in this region  likely originates from a PWN powered by the bright pulsar PSR J1838-0655.

The \gray\ emission from the northwest part (region A) can be modelled by an ellipse with the semimajor and semiminor axis of $0.5^{\circ}$ and $0.25^{\circ}$, respectively. The GeV emission has a hard spectrum with a photon index of about $2$ and is partially coincide with the TeV source MAGIC J1835-069.  The possible origin of the \gray\ emission in this region  is the interaction of the CRs accelerated by SNR G24.7+0.6  with the surrounding gas clouds. 
%We find a hard photon index of $\sim 2$ in both the vicinities of the SNR G24.7+0.6 and RSGC 1 (region A and B) between the energy range of 1 GeV - 500 GeV, which is consistent with that of Westerlund 2.

The GeV \gray\ emission from the southeast region (region B) can be modeled as an ellipse with the semimajor and semiminor axis of $0.9^{\circ}$ and $0.5^{\circ}$, respectively,  and also reveal a hard \gray\ spectrum, without any hint of cutoff. 
The similar spatial and spectral property of this source make it a candidate of the clone of Cygnus Cocoon \citep{Ackermann11,Aharonian19}. Thus we argue that the most probable origin is the interaction of the  accelerated protons in the young massive star cluster RSGC 1  with ambient gas clouds, and the total CR proton energy is estimated to be as high as  $\sim 1\times10^{50}\ \rm erg$. 

In the energy range of Fermi LAT, there is no evidence of spectral cutoff, which set a lower limit of the spectral cutoff of the parent protons to be $\sim 10 ~\rm TeV$.  An interesting question is whether this source can be a proton PeVatron. Indeed, \citet{bykov14} and \citet{bykov15} have already suggested that the supernova shock colliding with a fast wind from massive stars provide the most efficient acceleration site and the accelerated proton energies can be as high as dozens of PeV. The further investigation of such a scenario require the \gray\ observations at higher energies. Up to now, there is no TeV counterpart of this source, but it should be noted  that this may be caused by the limited field of view of current imaging air Cherenkov telescope arrays (ICATs) and the large size of this source, which make it difficult to find the off-regions in the standard data analysis of ICATs. In this regard, LHAASO \citep{lhaaso} may be the ideal instrument to investigate the TeV region of these objects. We plot the SED of region B and the LHAASO sensitivity \citep{lhaaso} in Fig.~\ref{fig:lhaaso}. It is evident that LHAASO has the ability to detect this source and set decisive  constraint on the cutoff energy in the near future, which would be crucial for understanding the acceleration mechanism in such system.

\section*{Acknowledgements}

This work is supported by the NSFC under grants 11421303, 11625312 and 11851304 and the National Key R\&D program of China under the grant 2018YFA0404203. Ruizhi Yang is supported  by the national youth thousand talents program in China.

\bibliographystyle{mnras}
%\bibstyle{aa}
%\bibliographystyle{plain}
\bibliography{ms}

\begin{thebibliography}{}
\makeatletter
\relax
\def\mn@urlcharsother{\let\do\@makeother \do\$\do\&\do\#\do\^\do\_\do\%\do\~}
\def\mn@doi{\begingroup\mn@urlcharsother \@ifnextchar [ {\mn@doi@}
  {\mn@doi@[]}}
\def\mn@doi@[#1]#2{\def\@tempa{#1}\ifx\@tempa\@empty \href
  {http://dx.doi.org/#2} {doi:#2}\else \href {http://dx.doi.org/#2} {#1}\fi
  \endgroup}
\def\mn@eprint#1#2{\mn@eprint@#1:#2::\@nil}
\def\mn@eprint@arXiv#1{\href {http://arxiv.org/abs/#1} {{\tt arXiv:#1}}}
\def\mn@eprint@dblp#1{\href {http://dblp.uni-trier.de/rec/bibtex/#1.xml}
  {dblp:#1}}
\def\mn@eprint@#1:#2:#3:#4\@nil{\def\@tempa {#1}\def\@tempb {#2}\def\@tempc
  {#3}\ifx \@tempc \@empty \let \@tempc \@tempb \let \@tempb \@tempa \fi \ifx
  \@tempb \@empty \def\@tempb {arXiv}\fi \@ifundefined
  {mn@eprint@\@tempb}{\@tempb:\@tempc}{\expandafter \expandafter \csname
  mn@eprint@\@tempb\endcsname \expandafter{\@tempc}}}

\bibitem[\protect\citeauthoryear{{Abdollahi} et~al.,}{{Abdollahi}
  et~al.}{2020}]{Fermi19}
{Abdollahi} S.,  et~al., 2020, \mn@doi [\apjs] {10.3847/1538-4365/ab6bcb},
  \href {https://ui.adsabs.harvard.edu/abs/2020ApJS..247...33A} {247, 33}

\bibitem[\protect\citeauthoryear{{Abramowski} et~al.,}{{Abramowski}
  et~al.}{2012}]{Abramowski12}
{Abramowski} A.,  et~al., 2012, \mn@doi [\aap] {10.1051/0004-6361/201117928},
  \href {https://ui.adsabs.harvard.edu/abs/2012A%26A...537A.114A} {537, A114}

\bibitem[\protect\citeauthoryear{{Ackermann} et~al.,}{{Ackermann}
  et~al.}{2011}]{Ackermann11}
{Ackermann} M.,  et~al., 2011, \mn@doi [Science] {10.1126/science.1210311},
  \href {https://ui.adsabs.harvard.edu/abs/2011Sci...334.1103A} {334, 1103}

\bibitem[\protect\citeauthoryear{{Aguilar} et~al.,}{{Aguilar}
  et~al.}{2015}]{ams02}
{Aguilar} M.,  et~al., 2015, \mn@doi [\prl] {10.1103/PhysRevLett.114.171103},
  \href {https://ui.adsabs.harvard.edu/abs/2015PhRvL.114q1103A} {114, 171103}

\bibitem[\protect\citeauthoryear{{Aharonian} et~al.,}{{Aharonian}
  et~al.}{2005}]{Aharonian05}
{Aharonian} F.,  et~al., 2005, \mn@doi [Science] {10.1126/science.1108643},
  \href {https://ui.adsabs.harvard.edu/abs/2005Sci...307.1938A} {307, 1938}

\bibitem[\protect\citeauthoryear{{Aharonian} et~al.,}{{Aharonian}
  et~al.}{2006}]{Aharonian06}
{Aharonian} F.,  et~al., 2006, \mn@doi [\apj] {10.1086/498013}, \href
  {https://ui.adsabs.harvard.edu/abs/2006ApJ...636..777A} {636, 777}

\bibitem[\protect\citeauthoryear{{Aharonian}, {Yang}  \& {de O{\~n}a
  Wilhelmi}}{{Aharonian} et~al.}{2019}]{Aharonian19}
{Aharonian} F.,  {Yang} R.,   {de O{\~n}a Wilhelmi} E.,  2019, \mn@doi [Nature
  Astronomy] {10.1038/s41550-019-0724-0}, \href
  {https://ui.adsabs.harvard.edu/abs/2019NatAs...3..561A} {3, 561}

\bibitem[\protect\citeauthoryear{{Bai} et~al.,}{{Bai} et~al.}{2019}]{lhaaso}
{Bai} X.,  et~al., 2019, arXiv e-prints, \href
  {https://ui.adsabs.harvard.edu/abs/2019arXiv190502773B} {p. arXiv:1905.02773}

\bibitem[\protect\citeauthoryear{{Blasi}}{{Blasi}}{2013}]{Blasi13}
{Blasi} P.,  2013, \mn@doi [\aapr] {10.1007/s00159-013-0070-7}, \href
  {https://ui.adsabs.harvard.edu/abs/2013A%26ARv..21...70B} {21, 70}

\bibitem[\protect\citeauthoryear{{Bolatto}, {Wolfire}  \& {Leroy}}{{Bolatto}
  et~al.}{2013}]{Bolatto13}
{Bolatto} A.~D.,  {Wolfire} M.,   {Leroy} A.~K.,  2013, \mn@doi [\araa]
  {10.1146/annurev-astro-082812-140944}, \href
  {https://ui.adsabs.harvard.edu/abs/2013ARA%26A..51..207B} {51, 207}

\bibitem[\protect\citeauthoryear{{Bykov}}{{Bykov}}{2014}]{bykov14}
{Bykov} A.~M.,  2014, \mn@doi [\aapr] {10.1007/s00159-014-0077-8}, \href
  {https://ui.adsabs.harvard.edu/abs/2014A&ARv..22...77B} {22, 77}

\bibitem[\protect\citeauthoryear{{Bykov}, {Ellison}, {Gladilin}  \&
  {Osipov}}{{Bykov} et~al.}{2015}]{bykov15}
{Bykov} A.~M.,  {Ellison} D.~C.,  {Gladilin} P.~E.,   {Osipov} S.~M.,  2015,
  \mn@doi [\mnras] {10.1093/mnras/stv1606}, \href
  {https://ui.adsabs.harvard.edu/abs/2015MNRAS.453..113B} {453, 113}

\bibitem[\protect\citeauthoryear{{Dame}, {Hartmann}  \& {Thaddeus}}{{Dame}
  et~al.}{2001}]{Dame01}
{Dame} T.~M.,  {Hartmann} D.,   {Thaddeus} P.,  2001, \mn@doi [\apj]
  {10.1086/318388}, \href
  {https://ui.adsabs.harvard.edu/abs/2001ApJ...547..792D} {547, 792}

\bibitem[\protect\citeauthoryear{{Davies}, {Figer}, {Law}, {Kudritzki},
  {Najarro}, {Herrero}  \& {MacKenty}}{{Davies} et~al.}{2008}]{Davies08}
{Davies} B.,  {Figer} D.~F.,  {Law} C.~J.,  {Kudritzki} R.-P.,  {Najarro} F.,
  {Herrero} A.,   {MacKenty} J.~W.,  2008, \mn@doi [\apj] {10.1086/527350},
  \href {https://ui.adsabs.harvard.edu/abs/2008ApJ...676.1016D} {676, 1016}

\bibitem[\protect\citeauthoryear{{Davies}, {de La Fuente}, {Najarro}, {Hinton},
  {Trombley}, {Figer}  \& {Puga}}{{Davies} et~al.}{2012}]{davies12}
{Davies} B.,  {de La Fuente} D.,  {Najarro} F.,  {Hinton} J.~A.,  {Trombley}
  C.,  {Figer} D.~F.,   {Puga} E.,  2012, \mn@doi [\mnras]
  {10.1111/j.1365-2966.2011.19840.x}, \href
  {https://ui.adsabs.harvard.edu/abs/2012MNRAS.419.1860D} {419, 1860}

\bibitem[\protect\citeauthoryear{{Drury}}{{Drury}}{2012}]{Drury12}
{Drury} L.~O.~.,  2012, \mn@doi [Astroparticle Physics]
  {10.1016/j.astropartphys.2012.02.006}, \href
  {https://ui.adsabs.harvard.edu/abs/2012APh....39...52D} {39, 52}

\bibitem[\protect\citeauthoryear{{Figer}, {MacKenty}, {Robberto}, {Smith},
  {Najarro}, {Kudritzki}  \& {Herrero}}{{Figer} et~al.}{2006}]{Figer06}
{Figer} D.~F.,  {MacKenty} J.~W.,  {Robberto} M.,  {Smith} K.,  {Najarro} F.,
  {Kudritzki} R.~P.,   {Herrero} A.,  2006, \mn@doi [\apj] {10.1086/503275},
  \href {https://ui.adsabs.harvard.edu/abs/2006ApJ...643.1166F} {643, 1166}

\bibitem[\protect\citeauthoryear{{Finkbeiner}}{{Finkbeiner}}{2003}]{Finkbeiner03}
{Finkbeiner} D.~P.,  2003, \mn@doi [\apjs] {10.1086/374411}, \href
  {https://ui.adsabs.harvard.edu/abs/2003ApJS..146..407F} {146, 407}

\bibitem[\protect\citeauthoryear{{Foreman-Mackey}, {Hogg}, {Lang}  \&
  {Goodman}}{{Foreman-Mackey} et~al.}{2013}]{Foreman13}
{Foreman-Mackey} D.,  {Hogg} D.~W.,  {Lang} D.,   {Goodman} J.,  2013, \mn@doi
  [\pasp] {10.1086/670067}, \href
  {http://adsabs.harvard.edu/abs/2013PASP..125..306F} {125, 306}

\bibitem[\protect\citeauthoryear{{Froebrich} \& {Scholz}}{{Froebrich} \&
  {Scholz}}{2013}]{froebrich13}
{Froebrich} D.,  {Scholz} A.,  2013, \mn@doi [\mnras] {10.1093/mnras/stt1633},
  \href {https://ui.adsabs.harvard.edu/abs/2013MNRAS.436.1116F} {436, 1116}

\bibitem[\protect\citeauthoryear{{Fujita}, {Nakanishi}, {Muller}, {Kobayashi},
  {Saito}, {Yasui}, {Kikuchi}  \& {Yoshinaga}}{{Fujita}
  et~al.}{2014}]{Fujita14}
{Fujita} Y.,  {Nakanishi} H.,  {Muller} E.,  {Kobayashi} N.,  {Saito} M.,
  {Yasui} C.,  {Kikuchi} H.,   {Yoshinaga} K.,  2014, \mn@doi [\pasj]
  {10.1093/pasj/pst020}, \href
  {https://ui.adsabs.harvard.edu/abs/2014PASJ...66...19F} {66, 19}

\bibitem[\protect\citeauthoryear{{Gotthelf} \& {Halpern}}{{Gotthelf} \&
  {Halpern}}{2008}]{Gotthelf08}
{Gotthelf} E.~V.,  {Halpern} J.~P.,  2008, \mn@doi [\apj] {10.1086/588779},
  \href {https://ui.adsabs.harvard.edu/abs/2008ApJ...681..515G} {681, 515}

\bibitem[\protect\citeauthoryear{{H.E.S.S.~Collaboration}
  et~al.,}{{H.E.S.S.~Collaboration} et~al.}{2015}]{Abramowski15}
{H.E.S.S.~Collaboration} et~al., 2015, \mn@doi [Science]
  {10.1126/science.1261313}, \href
  {https://ui.adsabs.harvard.edu/abs/2015Sci...347..406H} {347, 406}

\bibitem[\protect\citeauthoryear{{HI4PI Collaboration} et~al.,}{{HI4PI
  Collaboration} et~al.}{2016}]{HI4PI16}
{HI4PI Collaboration} et~al., 2016, \mn@doi [\aap]
  {10.1051/0004-6361/201629178}, \href
  {https://ui.adsabs.harvard.edu/abs/2016A%26A...594A.116H} {594, A116}

\bibitem[\protect\citeauthoryear{{Katsuta}, {Uchiyama}  \& {Funk}}{{Katsuta}
  et~al.}{2017}]{Katsuta17}
{Katsuta} J.,  {Uchiyama} Y.,   {Funk} S.,  2017, \mn@doi [\apj]
  {10.3847/1538-4357/aa6aa3}, \href
  {https://ui.adsabs.harvard.edu/abs/2017ApJ...839..129K} {839, 129}

\bibitem[\protect\citeauthoryear{{Lande} et~al.,}{{Lande}
  et~al.}{2012}]{Lande12}
{Lande} J.,  et~al., 2012, \mn@doi [\apj] {10.1088/0004-637X/756/1/5}, \href
  {https://ui.adsabs.harvard.edu/abs/2012ApJ...756....5L} {756, 5}

\bibitem[\protect\citeauthoryear{{Leahy}}{{Leahy}}{1989}]{Leahy89}
{Leahy} D.~A.,  1989, \aap, \href
  {https://ui.adsabs.harvard.edu/abs/1989A%26A...216..193L} {216, 193}

\bibitem[\protect\citeauthoryear{{Lebrun} et~al.,}{{Lebrun}
  et~al.}{1983}]{Lebrun83}
{Lebrun} F.,  et~al., 1983, \mn@doi [\apj] {10.1086/161440}, \href
  {https://ui.adsabs.harvard.edu/abs/1983ApJ...274..231L} {274, 231}

\bibitem[\protect\citeauthoryear{{MAGIC Collaboration} et~al.,}{{MAGIC
  Collaboration} et~al.}{2019}]{MAGIC19}
{MAGIC Collaboration} et~al., 2019, \mn@doi [\mnras] {10.1093/mnras/sty3387},
  \href {https://ui.adsabs.harvard.edu/abs/2019MNRAS.483.4578M} {483, 4578}

\bibitem[\protect\citeauthoryear{{Petriella}, {Paron}  \&
  {Giacani}}{{Petriella} et~al.}{2008}]{Petriella08}
{Petriella} A.,  {Paron} S.,   {Giacani} E.,  2008, Boletin de la Asociacion
  Argentina de Astronomia La Plata Argentina, \href
  {https://ui.adsabs.harvard.edu/abs/2008BAAA...51..209P} {51, 209}

\bibitem[\protect\citeauthoryear{{Petriella}, {Paron}  \&
  {Giacani}}{{Petriella} et~al.}{2010}]{Petriella10}
{Petriella} A.,  {Paron} S.,   {Giacani} E.,  2010, Boletin de la Asociacion
  Argentina de Astronomia La Plata Argentina, \href
  {https://ui.adsabs.harvard.edu/abs/2010BAAA...53..221P} {53, 221}

\bibitem[\protect\citeauthoryear{{Planck Collaboration} et~al.,}{{Planck
  Collaboration} et~al.}{2016}]{Planck16}
{Planck Collaboration} et~al., 2016, \mn@doi [\aap]
  {10.1051/0004-6361/201525967}, \href
  {https://ui.adsabs.harvard.edu/abs/2016A%26A...594A..10P} {594, A10}

\bibitem[\protect\citeauthoryear{{Popescu}, {Yang}, {Tuffs}, {Natale},
  {Rushton}  \& {Aharonian}}{{Popescu} et~al.}{2017}]{Popescu17}
{Popescu} C.~C.,  {Yang} R.,  {Tuffs} R.~J.,  {Natale} G.,  {Rushton} M.,
  {Aharonian} F.,  2017, \mn@doi [\mnras] {10.1093/mnras/stx1282}, \href
  {http://adsabs.harvard.edu/abs/2017MNRAS.470.2539P} {470, 2539}

\bibitem[\protect\citeauthoryear{{Portegies Zwart}, {McMillan}  \&
  {Gieles}}{{Portegies Zwart} et~al.}{2010}]{zwart10}
{Portegies Zwart} S.~F.,  {McMillan} S.~L.~W.,   {Gieles} M.,  2010, \mn@doi
  [\araa] {10.1146/annurev-astro-081309-130834}, \href
  {https://ui.adsabs.harvard.edu/abs/2010ARA%26A..48..431P} {48, 431}

\bibitem[\protect\citeauthoryear{{Reich}, {Furst}  \& {Sofue}}{{Reich}
  et~al.}{1984}]{Reich84}
{Reich} W.,  {Furst} E.,   {Sofue} Y.,  1984, \aap, \href
  {https://ui.adsabs.harvard.edu/abs/1984A%26A...133L...4R} {133, L4}

\bibitem[\protect\citeauthoryear{{Sodroski}, {Odegard}, {Arendt}, {Dwek},
  {Weiland}, {Hauser}  \& {Kelsall}}{{Sodroski} et~al.}{1997}]{Sodroski97}
{Sodroski} T.~J.,  {Odegard} N.,  {Arendt} R.~G.,  {Dwek} E.,  {Weiland} J.~L.,
   {Hauser} M.~G.,   {Kelsall} T.,  1997, \mn@doi [\apj] {10.1086/303961},
  \href {https://ui.adsabs.harvard.edu/abs/1997ApJ...480..173S} {480, 173}

\bibitem[\protect\citeauthoryear{{Yang} \& {Aharonian}}{{Yang} \&
  {Aharonian}}{2017}]{Yang17}
{Yang} R.-z.,  {Aharonian} F.,  2017, \mn@doi [\aap]
  {10.1051/0004-6361/201630213}, \href
  {https://ui.adsabs.harvard.edu/abs/2017A%26A...600A.107Y} {600, A107}

\bibitem[\protect\citeauthoryear{{Yang}, {de O{\~n}a Wilhelmi}  \&
  {Aharonian}}{{Yang} et~al.}{2018}]{Yang18}
{Yang} R.-z.,  {de O{\~n}a Wilhelmi} E.,   {Aharonian} F.,  2018, \mn@doi
  [\aap] {10.1051/0004-6361/201732045}, \href
  {https://ui.adsabs.harvard.edu/abs/2018A%26A...611A..77Y} {611, A77}

\bibitem[\protect\citeauthoryear{{Zabalza}}{{Zabalza}}{2015}]{naima}
{Zabalza} V.,  2015, in 34th International Cosmic Ray Conference (ICRC2015).
  p.~922 (\mn@eprint {arXiv} {1509.03319})

\makeatother
\end{thebibliography}

% Don't change these lines
\bsp	% typesetting comment
\label{lastpage}
\end{document}